\documentclass[12pt,numbers]{article}

\usepackage[margin=2.5cm]{geometry}
\usepackage{soul}
\usepackage{braket, siunitx}
\usepackage{graphicx}
\usepackage{dcolumn}
\usepackage{bm}
\usepackage{braket, siunitx}
\usepackage{amssymb, amsmath}
\usepackage{multirow}
\usepackage[acronym]{glossaries}
\usepackage[autostyle]{csquotes}
\usepackage{algorithm}
\usepackage{algpseudocode}
\usepackage{setspace}
\usepackage[sort&compress,numbers]{natbib}
\usepackage{authblk}
\usepackage{xcolor}
\usepackage{booktabs}
\usepackage{tablefootnote}
\usepackage{footnote}
\usepackage{float}

\usepackage{doi}

\makeatletter
\def\@fnsymbol#1{\ensuremath{\ifcase#1\or \ddagger\or z\or \ddagger\or
\mathsection\or \mathparagraph\or \|\or **\or \dagger\dagger
\or \ddagger\ddagger \else\@ctrerr\fi}}
\makeatother

\DeclareMathOperator*{\atanh}{atanh}

\newcommand*{\od}[3][]{\relax\ifmmode{\dfrac{\mathrm{d}^{#1} #2}{\mathrm{d} #3^{#1}}}\else{${\mathrm{d}^{#1} #2}/{\mathrm{d} #3^{#1}}$}\fi}
\newcommand*{\pd}[3][]{\relax\ifmmode{\dfrac{\partial^{#1} #2}{\partial #3^{#1}}}\else{${\partial^{#1} #2}/{\partial #3^{#1}}$}\fi}
\newcommand*{\eff}[0]{\mathrm{eff}}

\title{Bayesian Analysis of Interpretable Aging across Thousands of Lithium-ion Battery Cycles}

\author{Marc D. Berliner$^{a,1}$, Minsu Kim$^{a,1}$, Xiao Cui$^{b}$, Vivek N. Lam$^{b}$, Patrick~A.~Asinger$^{a}$, Martin Z. Bazant$^{a,c}$,  William C. Chueh$^{b}$, \newline Richard D. Braatz$^{a,}$\thanks{Corresponding author: Department of Chemical Engineering, Massachusetts Institute of Technology, 77 Massachusetts Avenue, Room E19-551, Cambridge, MA 02139, \texttt{braatz@mit.edu}.} \\
\small $^{a}$Department of Chemical Engineering, Massachusetts Institute of Technology, Cambridge, Massachusetts 02139, United States of America \\
\small $^{b}$Department of Materials Science and Engineering, Stanford University, Stanford, California 94305, United States of America \\
\small $^{c}$Department of Mathematics, Massachusetts Institute of Technology, Cambridge, Massachusetts 02139, United States of America}

\begin{document}

\maketitle

\footnotetext[1]{These authors contributed equally to this work.}

\begin{abstract}
The Doyle-Fuller-Newman (DFN) model is a common mechanistic model for lithium-ion batteries. The reaction rate constant and diffusivity within the DFN model are key parameters that directly affect the movement of lithium ions, thereby offering explanations for cell aging. This work investigates the ability to uniquely estimate each electrode's diffusion coefficients and reaction rate constants of 95 Tesla Model 3 cells with a nickel cobalt aluminum oxide (NCA) cathode and silicon oxide--graphite (LiC$_\text{6}$--SiO$_{\text{x}}$) anode. The parameters are estimated at intermittent diagnostic cycles over the lifetime of each cell. The four parameters are estimated using Markov chain Monte Carlo (MCMC) for uncertainty quantification (UQ) for a total of 7776 cycles at discharge C-rates of C/5, 1C, and 2C. While one or more anode parameters are uniquely identifiable over every cell's lifetime, cathode parameters become identifiable at mid- to end-of-life, indicating measurable resistive growth in the cathode. The contribution of key parameters to the state of health (SOH) is expressed as a power law. This model for SOH shows a high consistency with the MCMC results performed over the overall lifespan of each cell. Our approach suggests that effective diagnosis of aging can be achieved by predicting the trajectories of the parameters contributing to cell aging.
As such, extending our analysis with more physically accurate models building on DFN may lead to more identifiable parameters and further improved aging predictions.

\end{abstract}

\newpage
\doublespacing
\section{Introduction}

Lithium-ion batteries have become ubiquitous in modern technology, including in personal consumer devices and automobiles. Batteries are commonly modeled using porous electrode theory (PET) \cite{newman1975porous, doyle1993modeling, fuller1994relaxation, fuller1994simulation}, which includes electrochemical kinetics at the solid-electrolyte interfaces in the porous electrodes, transport through the electrolyte and in the solid particles, and thermodynamics modeled by as function of concentration for the open-circuit voltage (OCV). The Doyle-Fuller-Newman (DFN) model (also called the PET model and the Pseudo-Two-Dimensional (P2D) model) has one dimension as the position between the two metal contact points on the opposite sides of the electrode-separator-electrode sandwich and the second dimension as the distance from the center of a solid particle. The common practice is to fit a half dozen effective transport and kinetic coefficients in the DFN model to battery cycling data \cite{doyle1993modeling}, and then to use the model to explore changes in the battery design or the operating conditions.

When fitting model parameters to experimental data, an important consideration is whether the data contain sufficient information to uniquely specify the model parameters. Answering this question is referred to as an \emph{identifiability analysis}. While 
past studies have employed structural and linearized identifiability analyses to show that multiple combinations of effective transport and kinetic coefficients can produce nearly identical voltage discharge curves, more recent articles take an alternative approach of carrying out a fully nonlinear quantitative identifiability analysis for the DFN model \cite{galuppini2023nonlinear, berliner2021nonlinear}. The recent approach provides precise information on the uncertainty of the combinations of the estimated model parameters and makes precise statements as to which of the original model parameters are unlikely to be practically identifiable for commercial lithium-ion batteries.

Numerous publications have fit various lithium-ion battery parameters to cycling data using different models and methods.
For example, a local linearized sensitivity analysis was applied to separate groups of thermodynamic and kinetic DFN model parameters regressed on low-current discharge curves and current pulses, respectively \cite{jin2018parameter}.
Jokar et al.\ \cite{jokar2016inverse} implemented a local linearized sensitivity analysis, which was coupled with a genetic algorithm applied to a simplified P2D model to establish the time periods where parameters greatly affect a discharge voltage curve under high and low C-rates.
Another study \cite{lopez2016computational} applied linearized local sensitivities and a Monte Carlo (MC)-based covariance analysis to study identifiability from discharge curves and the electrolyte concentration in the center of the separator.
A fit of a hundred DFN model parameters to experimental battery cycling data using a genetic algorithm \cite{forman2012genetic} found that only a small subset of the parameters was identifiable.
The identifiability of parameters in the single-particle (SP) model with electrolyte dynamics under a constant state of charge (SOC), represented in terms of probability density functions (PDFs) of model parameters, has been quantified by Markov chain Monte Carlo (MCMC) \cite{aitio2020bayesian}. Unidentifiability in the solid-state diffusion parameters was alleviated by applying a sinusoidal pulse to achieve a range of SOC. At low C-rates, the SP model has a comparable error to the P2D model \cite{kemper2015simplification}.
The MCMC method has also been applied to quantify uncertainties in five effective transport and kinetic coefficients in the DFN model \cite{berliner2021nonlinear, ramadesigan2011parameter}. Of these five parameters, only the anode solid diffusion coefficient was found to be globally identifiable \cite{berliner2021nonlinear}. Reductions in the effective transport and kinetic coefficients over the cycle life of a Li-ion battery were plotted and shown to follow a power law \cite{ramadesigan2011parameter}. The approach predicted voltage discharge profiles at future cycles from experimental data collected for the first 200 cycles.

Our previous work \cite{berliner2021nonlinear, galuppini2023nonlinear} applied a nonlinear identifiability analysis to electrochemical battery models based on PET and multiphase PET (MPET) \cite{smith2017multiphase}, respectively. Ref.\ \cite{berliner2021nonlinear} estimated five diffusion and kinetic coefficients from synthetically generated data of a single discharge curve. Here we extend this work to a dataset with silicon oxide--graphite/nickel cobalt aluminium lithium-ion (NCA/LiC$_6$--SiO$_{\text{x}}$) cells from a disassembled Tesla Model 3. The \textit{a posteriori} parameter distributions are estimated at every diagnostic cycle using MCMC.
First, MCMC methods are presented and shown to provide global nonlinear identifiability trends for a small subject of parameters.
Second, unknown physical properties of the NCA/LiC$_6$--SiO$_{\text{x}}$ cell, which are estimated experimentally by inverting the model, are discussed.
Third, the model specifications are established for the standardized cycling conditions of the fleet of NCA cells.
Fourth, results are presented for the identifiability analysis and Bayesian estimation of each cycle with simple parameter fittings as a function of the State of Health (SOH).

This article is organized as follows: Section \ref{sec:background} provides theoretical background on the DFN model and Bayesian parameter identification. Section \ref{sec:physical_properties} describes the parameterization for explaining the cycling behavior of the NCA cell. Section \ref{sec:methodology} presents the overall scheme for parameter identification and provides details on the NCA cell data. Section \ref{result} represents the main results, which consist of identifiability and degradation diagnosis, and Section \ref{sec:conclusions} summarizes the article.

\section{Background}\label{sec:background}
\subsection{DFN model}\label{sec:PET}
The DFN model describes the microscopic physicochemical behavior of lithium-ion batteries and has been successfully applied to fast charging strategies \cite{kim2024fast, jiang2022fast}, aging analysis \cite{sordi2025degradation, liu2019aging}, lifetime prediction \cite{sulzer2021challenge}, and fault diagnosis \cite{mallarapu2020modeling}. Each porous electrode has an electrically conductive solid phase in close contact with a liquid electrolyte. The DFN has two dimensions: the $x$ direction, which moves across the length of the cell starting from the negative electrode through to the positive electrode, and the $r$ direction, the distance from the center of a porous electrode particle to its surface. The two porous electrodes are emersed in an electrolyte solution which conducts the flow of lithium ions. 

\begin{figure}
\centering
\includegraphics[width=0.7\textwidth]{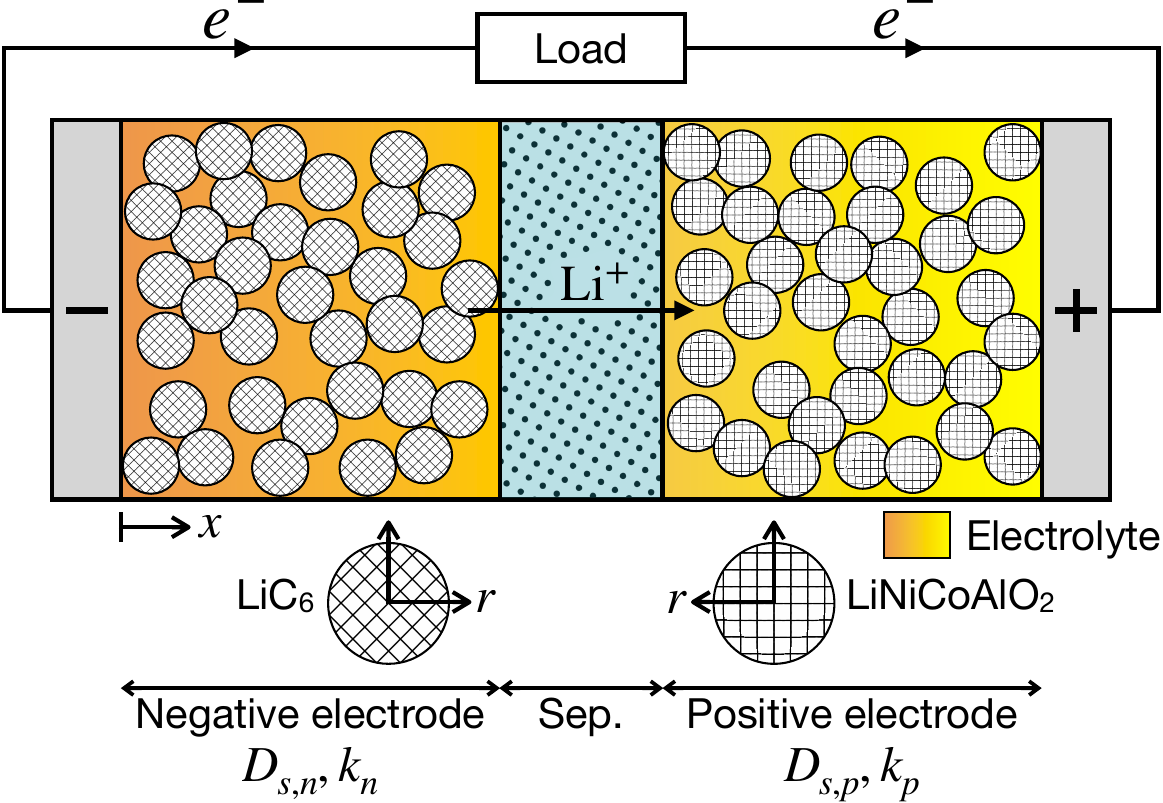}

\vspace{-0.2cm}

\caption{Schematic of the DFN model for an NCA/LiC$_6$--SiO$_{\text{x}}$ cell during discharge. The solid diffusion coefficients and reaction rate constants are listed under the sections whose physics they principally affect.}
\label{fig:PET}
\end{figure}

Solid-phase concentrations in each electrode control volume follow Fickian diffusion,
\begin{equation}\label{eqn:c_s_avg}
    \pd{c_s(x,r, t)}{t}=\frac{1}{r^{2}}  \frac{\partial}{\partial r}\!\left(r^2 D^{\eff}_i \pd{c_s(x,r, t)}{r}\right),
\end{equation}
with Neumann boundary conditions at the center and surface of the particle,
\begin{equation}\label{eqn:c_s_avg_bc}
    \left.\pd{c_s(x,r,t)}{r}\right|_{x,r=0}=0, \qquad \qquad\left. \pd{c_s(x,r, t)}{r}\right|_{r=R_p}=-\frac{j(x, t)}{D^{\eff}_{s,i}},
\end{equation}
where $j$ is the ionic flux in the electrodes, $D^{\eff}_{s, i}$ is the effective solid diffusion coefficient, and the superscript $i$ refers to a section of the cell: $\left\{ n, p \right\}$ are the positive and negative electrodes respectively, $s$ is the separator section, and $\left\{ a,z \right\}$ are the current collectors attached to the positive and negative electrodes respectively. The diffusion equation governs the electrolyte concentration,
\begin{equation}\label{eqn:c_e}
    \epsilon_i \pd{c_e(x, t)}{t} = \pd{}{x} \! \left(D^{\eff}_{i} \pd{c_e(x, t)}{x}\right)\!+a_i(1-t_+) j(x, t),
\end{equation}
where $\epsilon$ is porosity, $D^\eff$ is the effective electrolyte diffusion coefficient, $c_{e}$ is distribution of electrolyte concentration, $a$ is the surface area to volume ratio of the solid particles, and $t_+$ is the transference number. The boundary conditions at the electrode-current collector interfaces are
\begin{equation}\label{eqn:bc_bc1}
    \left. \pd{c_e(x, t)}{x} \right|_{x=0} = 0, \qquad\qquad \left. \pd{c_e(x, t)}{x} \right|_{x=L} = 0,
\end{equation}
and the interfacial flux terms are
\begin{equation}\label{eqn:c_e_bc2}
\begin{aligned}
    - D^{\eff}_n \left. \pd{c_e(x, t)}{x} \right|_{x=L_n^-} = &  - D^{\eff}_{s} \left. \pd{c_e(x, t)}{x} \right|_{x=L_n^+}, \\
    - D^{\eff}_s \left. \pd{c_e(x, t)}{x} \right|_{x=L_s^-} = &  - D^{\eff}_p \left. \pd{c_e(x, t)}{x} \right|_{x=L_s^+} ,
\end{aligned}
\end{equation}
where $L$ is the total length of the cell and $L_i$ is the distance measured from $x = 0$ to the end of section $i$.

Interfacial ionic fluxes couple the two phases in the electrode sections modeled by Butler-Volmer reaction kinetics, 
\begin{equation}\label{eqn:j}
    j(x, t) = 2 k^{\eff}_{i} \sqrt{c_e(x, t)\left(c^{\max}_{s, i}-c_{s}^{*}(x, t)\right) c_s^{*}(x, t)}  \, \sinh \!\left( \dfrac{0.5 F}{R T(x, t)} \eta(x, t)\right) ,
\end{equation}
where $k^\eff$ is the effective rate constant, $c^{\max}_{s}$ is the maximum solid concentration, $c_s^*$ is the solid particle surface concentration, $T$ is temperature, $\eta$ is the overpotential, $F$ is Faraday's constant, and $R$ is the ideal gas constant.

Partial differential equations for the solid and electrolyte potentials, $\Phi_s$ and $\Phi_e$ respectively, are
\begin{equation}\label{eqn:Phi_s}
    \pd{}{x} \! \left( \sigma^{\eff}_{i} \pd{\Phi_s (x,t)}{x} \right) = a_i F j(x,t),
\end{equation}
\begin{equation}\label{eqn:Phi_e}
    \pd{}{x} \! \left( \kappa^\eff(x,t) \pd{\Phi_e (x,t)}{x} \right) = - a_i F j(x,t) + \pd{}{x} \! \left(\dfrac{2 \kappa^\eff (x,t) R T(x, t)}{F}\left(1 - t_+\right) \pd{\ln c_e(x, t)}{x} \right) \!,
\end{equation}
where $\sigma^\eff$ and $\kappa^\eff$ are the effective solid and electrolyte conductivities, respectively. 
Boundary conditions for $\Phi_s$ incorporate the applied current $I(t)$,
\begin{equation}\label{eqn:phis_bc1}
    \left. \pd{\Phi_s(x, t)}{x} \right|_{x=0} = -I(t), \qquad\qquad \left. \pd{\Phi_s(x, t)}{x} \right|_{x=L} = -I(t),
\end{equation}
\begin{equation}\label{eqn:phis_bc2}
    \left. \pd{\Phi_s(x, t)}{x} \right|_{x=L_n} = 0, \qquad \qquad \left. \pd{\Phi_s(x, t)}{x} \right|_{x=L_s} = 0 .
\end{equation}
The boundary conditions for $\Phi_e$ mirror those for $c_e$,
\begin{equation}
\label{eqn:phie_bc1}
    \left. \Phi_e(x, t) \right|_{x=0} = 0, 
    \qquad\qquad \left. \pd{\Phi_e(x, t)}{x} \right|_{x=L} = 0,
\end{equation}
\begin{equation}
\label{eqn:phie_bc2}
\begin{aligned}
    - \kappa^{\eff}_{n} \left. \pd{\Phi_e(x, t)}{x} \right|_{x=L_n^-} = &  - \kappa^{\eff}_{s} \left. \pd{\Phi_e(x, t)}{x} \right|_{x=L_n^+}, \\
    - \kappa^{\eff}_{s} \left. \pd{\Phi_e(x, t)}{x} \right|_{x=L_s^-} = &  - \kappa^{\eff}_{p} \left. \pd{\Phi_e(x, t)}{x} \right|_{x=L_s^+} ,
\end{aligned}
\end{equation}
with the reference potential grounded to $0\,\text{V}$ at $x = 0$. For further details of the governing equations, see Ref.\ \cite{fuller1994relaxation}. We solve the equations using PETLION \cite{berliner2021petlion}, which is an efficient open-source DFN simulation tool in the Julia programming language. PETLION discretizes the PDEs in space using the Finite Volume Method (FVM) and solves forward in time using the Method of Lines (MoL) \cite{Schiesser}. The number of FVM control volumes ($N$) and solver tolerances ($\Delta_\textrm{abs}$: absolute value, $\Delta_\textrm{rel}$: relative value) affect the desired accuracy and computational complexity of the simulation. All solver settings are set to defaults ($N = 10$, $\Delta_\textrm{abs}$ = 10$^{-6}$, $\Delta_\textrm{rel}$ = 10$^{-3}$), as increasing the number of control volumes and decreasing the tolerance significantly increased the computational budget without meaningfully improving accuracy \cite{kim2024fast}.

\subsection{Bayesian Parameter identification and Identifiability}\label{sec:MCMC}

Parameter identification through Bayesian inference is a method for quantifying the uncertainty about estimated parameters \cite{golberg1989genetic, lee2024multi}. Bayesian inference, based on Bayes’ theorem, is explained by the relationship between the \textit{posterior} distribution $\text{P}(\theta|Y)$, \textit{prior} distribution $\text{P}(\theta)$, \textit{likelihood} $\text{P}(Y|\theta)$, and \textit{marginal likelihood} $\text{P}(Y)$,
\begin{equation}\label{eqn:bayes}
\text{P}( \theta|Y )=\frac{\text{P}( Y|\theta)\,\text{P}(\theta)}{\text{P}(Y)},
\end{equation}
where $Y$ is a vector of voltage measurements $y_{j}$. To estimate the most promising subset of parameters, the \textit{Maximum A Posteriori} (MAP) is obtained by maximizing the posterior probability,
\begin{equation}\label{eqn:MAP1}
\theta^{*}_{\text{MAP}}=\underset{\theta}{\text{argmax}}\ \text{P}( \theta|Y)=\underset{\theta}{\text{argmax}}\  \text{P}( Y|\theta)\text{P}(\theta),
\end{equation}
where is $\text{P}(Y)$ is a normalization constant which may be ignored when solving the maximization. In practice, it is common to instead minimize a negative log transformation of \eqref{eqn:MAP1},
\begin{equation}\label{eqn:MAP2}
\theta^{*}_{\text{MAP}} = \underset{\theta}{\text{argmin}}\  \ln \text{P}( Y|\theta) + \ln\text{P}(\theta),
\end{equation}
which gives the same parameter value. 
If the prior distribution is a uniform or normal distribution with an infinitely large variance, the influence of the prior distribution can be neglected and the equation becomes the \emph{Maximum Likelihood Estimate} (MLE),
\begin{equation}\label{eqn:MLE1}
\theta^{*}_{\text{MLE}}=\underset{\theta}{\text{argmin}}\  \ln \text{P}(Y|\theta).
\end{equation}
The MLE for a model with Gaussian error simplifies to the familiar sum of squared residuals,
\begin{equation}\label{eqn:MLE2}
\theta^{*}_{\text{MLE}}=\underset{\theta}{\text{argmin}} \sum_{j=1}^{N_d}\left( \frac{y_{j}-\widehat{y}_{j}(\theta)}{\sigma_{\epsilon}} \right)^{\!\!2}.
\end{equation}

The uncertainty associated with the estimated parameters are quantified through the confidence regions. The DFN model exhibits a highly nonlinear relationship between parameters and model predictions, which suggests that the confidence regions may not be a hyperellipsoid \cite{johnson2002applied}. Linearization approaches 
may produce highly inaccurate estimates for highly nonlinear systems \cite{berliner2021nonlinear, galuppini2023nonlinear}. 
Nonlinear systems can be handled by relating the optimal estimates of parameters derived from the log-likelihood function to the chi-squared distribution  \cite{meeker1995teaching}. 
With the definition
\begin{equation}\label{eqn:chi}
\chi^{2}(\theta)=\sum_{j=1}^{N_{d}}\left( \frac{y_{j}-\widehat{y}_{j}(\theta)}{\sigma_{\epsilon}} \right)^{\!\!2},
\end{equation}
the nonlinear confidence region $R_{\alpha}$ can be expressed as all $\theta$ that satisfy the inequality
\begin{equation}\label{eqn:CI}
R_{\alpha} = \left\{ \theta:\chi^{2}(\theta)-\chi^{2}(\theta^{*})\le \chi^{2}_{N_{p}}(1-\alpha)  \right\},
\end{equation}
where $\chi^2_{N_p}$ is the chi-squared distribution with $N_p$ degrees of freedom, $N_p$ is the number of parameters, and $\alpha$ is the significance level (e.g., an $\alpha$ of 0.01 corresponds to a 99\% confidence region).

 MCMC is an efficient approach for uncertainty evaluation for highly nonlinear systems. MCMC employs the Metropolis-Hastings algorithm which can sample from complex high-dimensional PDFs \cite{haario2006dram, roberts2009examples}. Parameters are initiated at an initial value, $\theta_0$, and random perturbations are introduced to these parameters \cite{chib1995understanding}. During each iteration $t$, an objective function $f(\theta)$, such as the sum of squared residuals, 
 is calculated. The suggested parameter $\theta'$ is then accepted or rejected based on an acceptance ratio, serving as the criterion for determining whether to adopt the next parameter set.

\section{Parameterization: Physical Properties of NCA 
Cell}\label{sec:physical_properties}

The first step to performing an identifiability analysis for the NCA cells is accurately modeling the cell. The properties of the electrolyte are estimated as a function of electrolyte concentration and temperature. The cells have a small amount of silicon oxide in the anode which differentiates them from conventional chemistries in the literature. To address this, the OCV functions for each electrode are regressed in cycling experiments of half-cells containing positive and negative electrodes with C/50 charge and C/60 discharge. Then, the half-cell OCVs are regressed against the full-cell OCV using nonlinear optimization to estimate the stoichiometry limits. The estimated functions for the properties of the electrolyte and electrode are presented in detail in Sections \ref{sec:electrolyte_fitting} and \ref{sec:electrode_fitting}, respectively.

The DFN model parameters used for the pristine cell simulation are listed in Table \ref{tab:tesla_param}. The reaction rate constants and diffusion coefficients at each electrode are estimated through identifiability analysis, and the other parameters are considered constants throughout the cell lifetime.

\begin{table}
\caption{Description of the parameter set of the NCA/LiC$_{6}$--Si cell}
\label{tab:tesla_param}
\centering
\resizebox{16cm}{!}{%
\begin{tabular}{cclccccc}
\hline
Parameter & Unit & \multicolumn{1}{c}{Description}               & NCA & Cathode & Separator & Anode & LiC$_6$--Si \\ \hline
$D_{s, i}$& m$^2$/s & Solid-phase diffusivity           &  -- &     8.716$\times$10$^{-14}$    &   --   &    1.018$\times$10$^{-13}$   &   --   \\
$k_{i}$& m$^{5/2}$/(mol$^{1/2}$s) & Reaction rate constant  & --  &   4.438$\times$10$^{-10}$      &     --    &   6.837$\times$10$^{-12}$    &  --  \\
$l_{i}$& m & Thickness                                           &  1.0$\times$10$^{-5}$  &  6.4$\times$10$^{-5}$  &  1.0$\times$10$^{-5}$   &  8.3$\times$10$^{-5}$  &   1.0$\times$10$^{-5}$   \\
$\epsilon_{i}$&   --   & Porosity                                & --  &    0.230     &    0.359       &   0.147    &   -- \\
$D_{i}$& m$^{2}$/s & Electrolyte diffusivity               &  --  &    5.0$\times$10$^{-10}$   &   5.0$\times$10$^{-10}$   &   5.0$\times$10$^{-10}$   &    --  \\
$R^p_i$& m & Particle radius                                 &  --  &    1.1$\times$10$^{-5}$     &        --   &    1.6$\times$10$^{-5}$   &    --  \\
$c^\mathrm{init}_{e, i}$& mol/m$^{3}$  & Initial concentration in the electrolyte &   --  &   1,200   &   1,200   &   1,200  &  --  \\
$c^{\max}_{s, i}$& mol/m$^{3}$  & Maximum solid-phase concentration &   --  &    54,422     &     --      &    28,967   &   --   \\
$\rho_{i}$& kg/m$^{3}$ & Density                                &  2,700   &    2,500     &    1,100   &    2,500   &  8,940 \\
$C_{p,i}$& J/(kg K) & Specific heat                   &  897  &   700      &   700    &   700   &   385   \\
$\lambda_{i}$&   W/(m K)   & Thermal conductivity   &  237   &     2.1    &   0.16     &    1.7   &  401  \\
$\sigma_{i}$& S/m & Solid-phase conductivity              &   3.55$\times$10$^{7}$  &    100     &    --       &    100   &   5.96$\times$10$^{7}$   \\
$\epsilon_{s, i}$& -- & Active material fraction                          &  --  &    0.745     &     --    &    0.828   &   --  \\
 Brugg & -- & Bruggeman coefficient                              &  -- &   1.5   &    1.5   &   1.5   &  --  \\
$t_{+}$&  0.455 & Transference number                            &  -- &   --    &     --    &   --  &   -- \\
$E^{D^{s}_{i}}_{a}$& J/mol & Solid-phase diffusion activation energy  &  --   &   5,000     &     --      &   5,000    &   --   \\
$E^{k_{i}}_{a}$& J/mol & Reaction constant activation energy      &   --  &     5,000    &     --     &   5,000    &   --   \\
$\Theta^{\max}_{i}$& -- & Maximum stoichiometry limits             &  --  &     0.160    &     --      &   0.923    &    --  \\
$\Theta^{\min}_{i}$& -- & Minimum stoichiometry limits             &  --   &    0.859     &     --      &   0.014    &    --  \\
$T_\mathrm{amb}$& 298.15 K & Ambient temperature        &  --   &      --   &     --      &    --   &   --   \\
$F$& 96485 C/mol & Faraday's constant        &  --   &      --   &     --      &    --   &   --   \\
$R$& 8.314472 J/(mol K)  & Universal gas constant        &  --   &      --   &     --      &    --   &   --   \\ \hline
\end{tabular}
}
\end{table}

\subsection{Electrolyte}\label{sec:electrolyte_fitting}
The electrolyte conductivity $\kappa$ and diffusivity $D$ are tabulated as functions of electrolyte concentration $c_e$ and temperature $T$ using the Advanced Electrolyte Model \cite{gering2017prediction} for an EC/EMC/DMC electrolyte mix. Empirical equations for $\kappa(c_e,T)$ and $D(c_e,T)$ are fit using response surface methodology \cite{khuri2010response}, 
\begin{equation}\label{eqn:kappa}
    \kappa(c_e, T) = \sum_{i=0}^3 \sum_{j=0}^1 a_{ij} ( c_e^i T^j),
\end{equation}
\begin{equation}\label{eqn:D}
    D(c_e, T) = \sum_{i=0}^4 \sum_{j=0}^2 a_{ij}( c_e^i  T^j).
\end{equation}

\begin{table}
\caption{Coefficient ($a_{ij}$) for electrolyte conductivity ($\kappa(c_e,T)$) in \eqref{eqn:kappa}.}
\label{tab:kappa}
\centering

\resizebox{7cm}{!}{%
\begin{tabular}{ccc}
\hline
     & $j=0$ & $j=1$ \\ \hline
$i=0$ &  $-$5.182$\times$10$^{-1}$   &  1.696$\times$10$^3$   \\
$i=1$ &  $-$6.518$\times$10$^{-3}$   &  3.034$\times$10$^{-5}$   \\
$i=2$ & 1.446$\times$10$^{-6}$  & $-$1.049$\times$10$^{-8}$    \\
$i=3$ &  3.047$\times$10$^{-10}$  &  0 \\ \hline
\end{tabular}
}
\end{table}

\begin{table}
\caption{Coefficient ($a_{ij}$) for electrolyte diffusion coefficients ($D(c_e,T)$) in \eqref{eqn:D}.}
\label{tab:D}
\centering
\begin{tabular}{cccc}
\hline
 & $j=0$ & $j=1$ & $j=2$ \\ \hline
 $i=0$&   1.864$\times$10$^{-8}$  &  $-$1.392$\times$10$^{-10}$   &   2.633$\times$10$^{-13}$  \\
 $i=1$&   0  &   3.133$\times$10$^{-14}$  &  $-$1.126$\times$10$^{-16}$   \\
 $i=2$&   0  &   $-$7.301$\times$10$^{-17}$  &  2.615$\times$10$^{-19}$   \\
 $i=3$&   0  &  5.120$\times$10$^{-20}$   &   $-$1.832$\times$10$^{-22}$  \\
 $i=4$&   0  &   $-$1.151$\times$10$^{-23}$  &  4.111$\times$10$^{-26}$   \\ \hline
\end{tabular}
\end{table}

\subsection{Electrode}\label{sec:electrode_fitting}
The anode is graphite doped with a silicon oxide, producing a bimodal particle radius distribution with peaks in different regions for graphite and silicon oxide particles \cite{moyassari2021impact, lee2008spherical}. That is, although even small amounts of silicon oxide dopant in graphite lead to significant structural differences, for simplicity of calculations, we assume the anode homogeneously, as a single particle with a radius of 16 $\mu$m.

The positive and negative OCVs are estimated with charge and discharge cycles at 0.1 and 0.2 mA, respectively (about C/60 and C/50).
The OCVs are fit with empirical equations as a function of solid lithium concentration,
\begin{equation}\label{eqn:OCV_n}
    U_n(\Theta_n) = a_0 + a_1 \exp\! \left( \dfrac{\Theta_n - b_1}{c_1} \right) \!+ \sum_{i=2}^4 a_i \tanh \!\left( \dfrac{\Theta_n - b_i}{c_i} \right),
\end{equation}
\begin{equation}\label{eqn:OCV_p}
    U_p(\Theta_p) = \sum_{i=1}^8 a_i \exp\! \left( -\left( \dfrac{\Theta_p - b_i}{c_i} \right)^{\!\!2} \right),
\end{equation}
\begin{equation}\label{eqn:normalized_c_s}
    \Theta_i = {c_{s,i}^*}/{c_{s,i}^{\max}},
\end{equation}
where $U_i$ is open-circuit voltage, $\Theta_i$ is the stoichiometry, and $c^{*}_{s,i}$ is solid-phase surface concentration. The coefficients for the empirical equations representing the open-circuit voltages of the negative electrode ($U_{n}$) and positive electrode ($U_{p}$) are detailed in Tables~\ref{tab:u_n} and~\ref{tab:u_p}, respectively.

\begin{table}
\caption{Coefficients ($a_i$, $b_i$, and $c_i$) for negative open-circuit voltage ($U_n$($\Theta_n$)) in \eqref{eqn:OCV_n}.}
\label{tab:u_n}
\centering
\begin{tabular}{cccc}
\hline
 & $a_{i}$ & $b_{i}$ & $c_{i}$ \\ \hline
 $i=0$ &  $-$48.99 & --  & --  \\
 $i=1$&  29.98 &  5.700$\times$10$^{-3}$ &  $-$5.093$\times$10$^{-2}$ \\
 $i=2$&  161.9 & $-$1.057$\times$10$^{-1}$  & 9.687$\times$10$^{-2}$  \\
 $i=3$&  $-$2.833$\times$10$^{-1}$ & 4.447$\times$10$^{-2}$  &  4.235$\times$10$^{-2}$ \\
 $i=4$& $-$47.77  & $-$18.95  &  7.041 \\
 $i=5$&  $-$65.06 &  2.268$\times$10$^{-3}$ &  1.160$\times$10$^{-3}$ \\ \hline
\end{tabular}
\end{table}

\begin{table}
\caption{Coefficients ($a_i$, $b_i$, and $c_i$) for positive open-circuit voltage ($U_p(\Theta_p)$ in \eqref{eqn:OCV_p}.}
\label{tab:u_p}
\centering
\begin{tabular}{cccc}
\hline
 & $a_i$ & $b_i$ & $c_i$ \\ \hline
$i=1$ & 1.456$\times$10$^{-1}$  &  7.961$\times$10$^{-1}$ & 6.035$\times$10$^{-2}$  \\
$i=2$& 4.205$\times$10$^{-1}$  &  9.489$\times$10$^{-1}$ &  4.229$\times$10$^{-2}$ \\
$i=3$&  1.008 &  6.463$\times$10$^{-1}$ & 1.034$\times$10$^{-1}$  \\
$i=4$&  1.350  &  7.378$\times$10$^{-1}$ & 9.513$\times$10$^{-2}$  \\
$i=5$& 2.526  & 2.953$\times$10$^{-1}$  &  2.019$\times$10$^{-1}$ \\
$i=6$&  2.636 & 5.372$\times$10$^{-1}$  & 1.758$\times$10$^{-1}$  \\
$i=7$&  3.285 &  8.922 &  1.414$\times$10$^{-1}$ \\
$i=8$& 172.1  & $-$1.344  & 7.371$\times$10$^{-1}$  \\ \hline
\end{tabular}
\end{table}

\section{Methodology}\label{sec:methodology}

\subsection{Bayesian Estimation and Identifiability Procedure}

The analysis of nonlinear identifiability has three steps: (1) estimating the posterior distribution of $\theta$ using the Metropolis-Hastings algorithm, (2) distinguishing between practically identifiable and unidentifiable parameters based on the probability densities, and (3) further categorizing the identifiable combinations through a gridded mesh showing the confidence regions. Once the parameter space is comprehensively mapped, a set of identifiable parameter groups is defined, which includes all identifiable and locally identifiable parameters.

The parameter space can be reduced by fixing unidentifiable parameters to an estimate or to physically meaningful upper and lower bounds. Likewise, equations considered insignificant because of the unidentifiable parameters can be excluded from the model to enhance computational efficiency. For instance, in the case of a very thin porous electrode that is not diffusion limited, the effects of diffusion within the porous electrode could be disregarded in the model. Alternatively, a less restrictive approach would involve incorporating prior values from the literature.

Practical identifiability is confirmed through the results obtained from Bayesian estimation. A parameter is deemed practically identifiable if a sufficiently large number of chains indicate that it is bounded. Conversely, if the chains include parameter values that can be arbitrarily large or small, then the parameter is practically unidentifiable; further investigation is needed to determine whether there are any identifiable combinations. All identifiable combinations encompass every practically identifiable parameter.

Locally identifiable parameters that contribute to \textit{identifiable combinations} are first evaluated using their probability densities. A parameter is likely part of the identifiable combinations if its probability density exhibits (1) a large peak and (2) a lower magnitude plateau at extreme values. The significant peak arises from identifiable combinations that include the parameter, while the plateau corresponds to the identifiable combinations that exclude it, indicating that the parameter is unidentifiable. Parameters with a uniform distribution do not belong to any identifiable combination sets. A detailed explanation of the methodology regarding identifiability, along with a simple example, is specified in previous work \cite{berliner2021nonlinear, galuppini2023nonlinear}.

\subsection{Model Specifications}\label{sec:model_specifications}

This article considers the key transport and kinetic parameters,  $\theta = [ D_{s,p}$, $D_{s,n}$, $k_p$, $k_n]^\top$, where $D_{s,p}$ and $D_{s,n}$ are the solid-phase diffusion coefficients of lithium in the cathode and anode respectively, and $k_p$ and $k_n$ are the electrochemical reaction rate constants for the cathode and anode respectively. These parameters lump the effects of multiple true material properties together \cite{newman1975porous}. The parameter identification is performed on a logarithmic basis of $\theta$, a standard approach to improve numerical convergence for parameters that can change by many orders of magnitude.

\begin{figure}[h]
\centering
\includegraphics[width=1\textwidth]{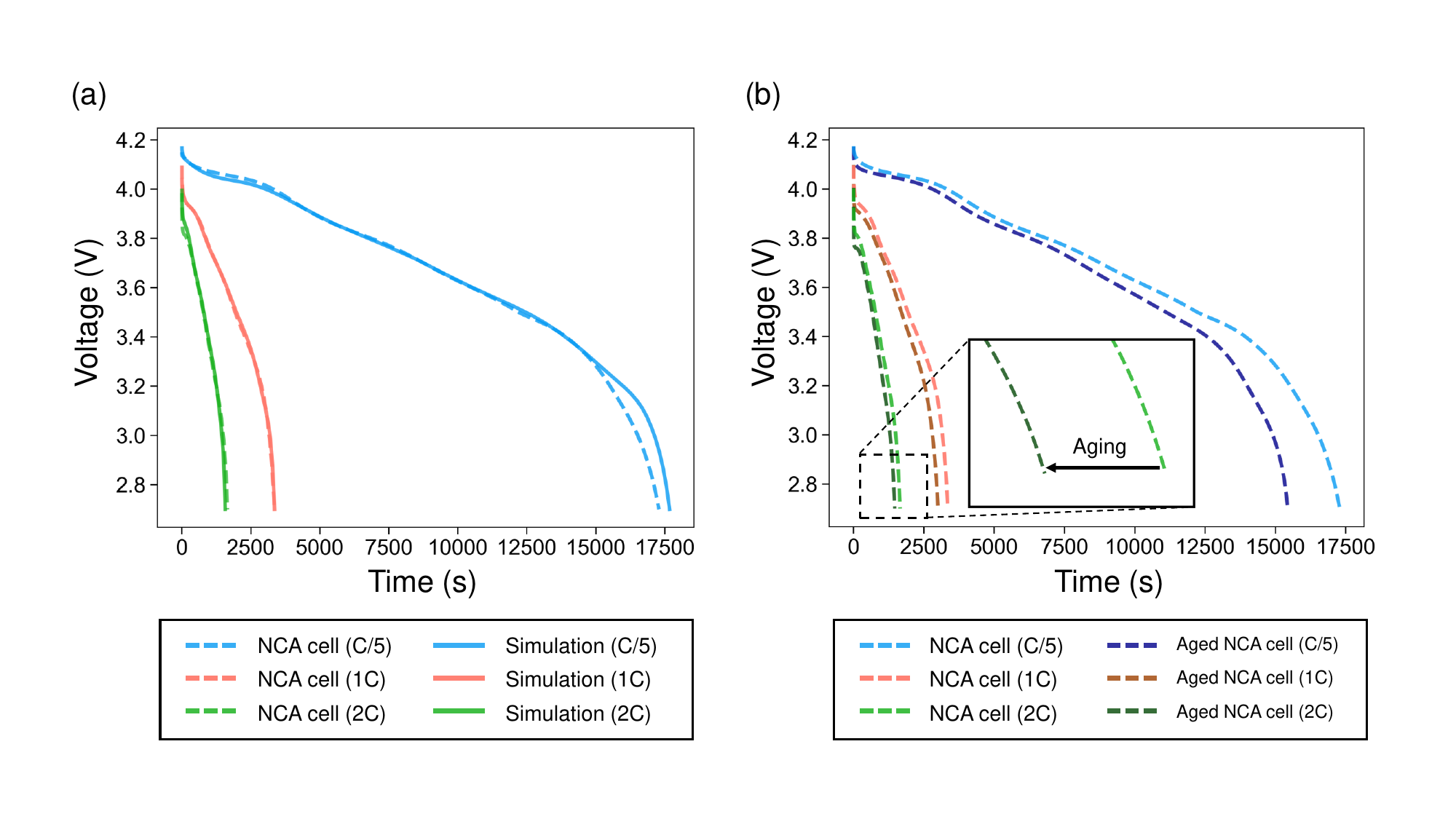}

\vspace{-1cm}

\caption{(a) Generation of high-fidelity battery models through parameterization and comparison with real experiments and  (b) comparison of cycling behavior of an aged cell and pristine cell.}
\label{fig:NCA cell data}
\end{figure}

The cycling data used for the identifiability analysis consisted of a representative sample of 95 cells, selected from a total of 363 cylindrical 21700 NCA cells extracted from a Tesla Model 3 provided by van Vlijmen et al.~\cite{van2023interpretable}. Diagnostic cycles consisting of a reset cycle, hybrid pulse power characterization (HPPC), and reference performance test (RPT) cycles at three discharge C-rates (C/5, 1C, and 2C) are performed every 100 cycles for the whole lifetime of the 95 cells to measure the capacity fade (see Ref.~\cite{van2023interpretable} for a detailed description of the NCA cell dataset). Only the RPT cycles among the diagnostic cycles are used for the identifiability analysis. The discharge voltage curve used for parameter identification is represented with high fidelity in battery simulations by applying the DFN model parameters identified in Section \ref{sec:physical_properties} (Fig.\ \ref{fig:NCA cell data}a). As cycling continues, the capacity irreversibly decreases (Fig.\ \ref{fig:NCA cell data}b). Four parameters ($D_{s,p}$, $D_{s,n}$, $k_p$, $k_n$) are estimated through the discharge curves from all RPTs of each cell, and the trajectory of each parameter is tracked as aging progresses. 

\section{Results and Discussion} \label{result}
\subsection{Parameter identifiability}
Visualizing the nonlinear confidence region can help interpret identifiability trends. A nonlinear confidence region depicts the error (e.g., the sum of squared residuals, chi-squared statistic, root-mean-squared error (RMSE)) as a function of the $N_p$-dimensional parameter space. Typically, the parameter space is gridded with sufficiently fine discretizations to show detailed resolution of the confidence region. Confidence regions of linear or linearized models depict hyperellipsoid confidence regions centered on $\theta^*$ \cite{beck1977parameter, berliner2021nonlinear}, but nonlinear confidence regions are not restricted to hyperellipsoid shapes. Highly nonlinear models (such as PET-based battery models) have been shown to exhibit \textit{banana-shaped} confidence regions in which an infinite number of parameter values give either the same or nearly the same quality of fit.

\begin{figure}[h]
\centering
\includegraphics[width=1\textwidth]{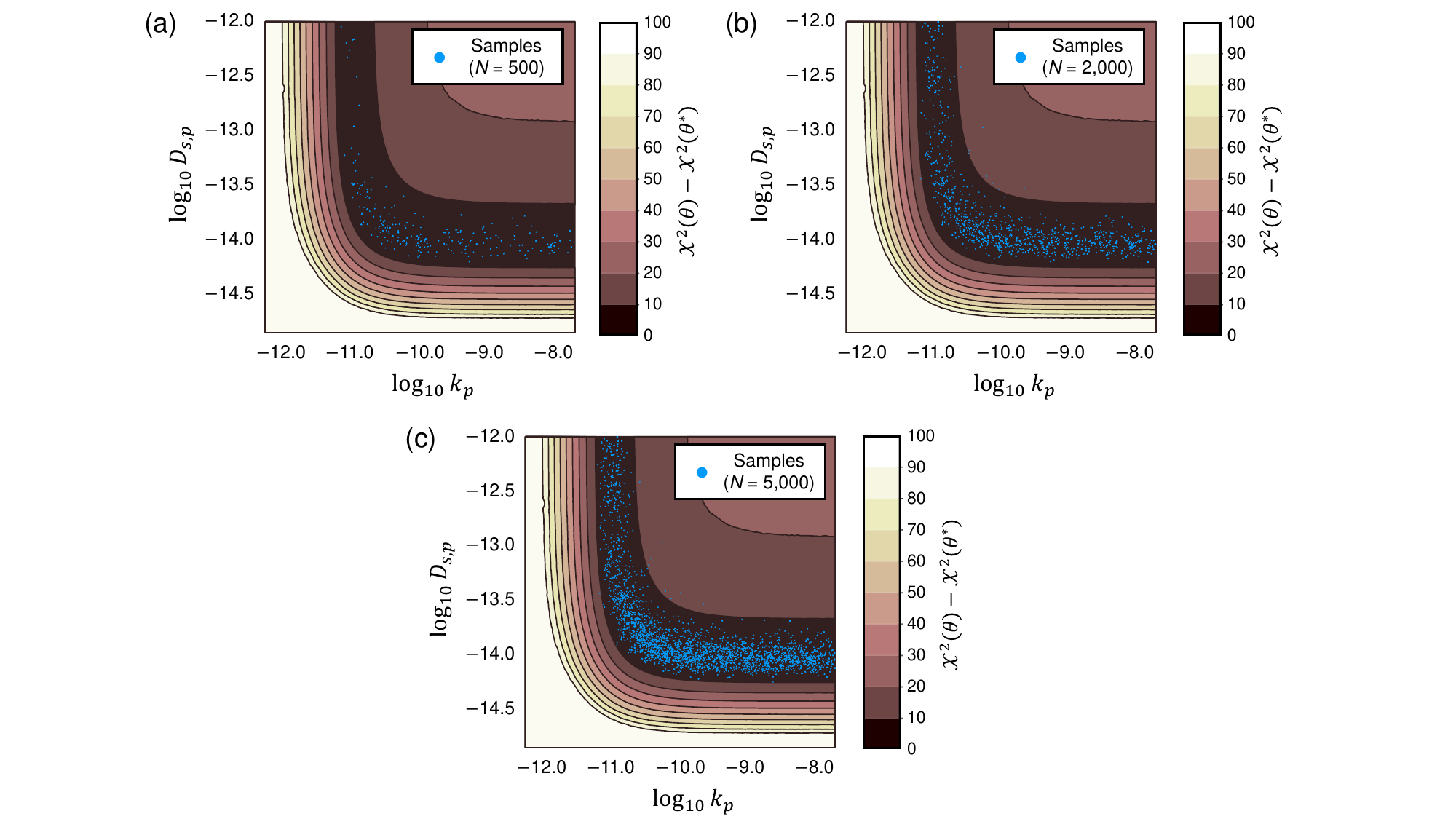}

\vspace{-0.3cm}

\caption{MCMC sampling of the 2-dimensional confidence region: (a) 500 samples, (b) 2000 samples, (c) 5000 samples.}
\label{fig:confidence_region}
\end{figure}

Fig.\ \ref{fig:confidence_region} shows an example confidence region as a function of the cathode diffusion coefficient $D_{s,p}$ and the rate constant $k_p$. Both $D_{s,p}$ and $k_p$ are \emph{practically unidentifiable} because the darkly shaded extends towards $+\infty$. The parameters appear to be locally identifiable in that a numerical optimization at any initial guess would converge to a point on the minimum curve that would locally appear to be minimum over $k_p$ for fixed $D_{s,p}$ and locally appear to be minimum over $D_{s,p}$ for fixed $k_p$. Inspection of the conference regions as a function of both parameters, as seen in Fig.\ \ref{fig:confidence_region}, shows that the two parameters are not globally identifiable. The two extremes of the minimum curve, where $k_p \to \infty$ and $D_{s,p} \to \infty$ respectively, show very different sensitivities on $k_p$ and $D_{s,p}$. Sensitivities of the parameter identification objective on the parameters can be very large or nearly zero depending on the value of the other parameter. This observation has strong parameter identification implications and implies that relying only on local sensitivities can lead to misleading results.

\begin{figure}[h]
\centering
\includegraphics[width=1\textwidth]{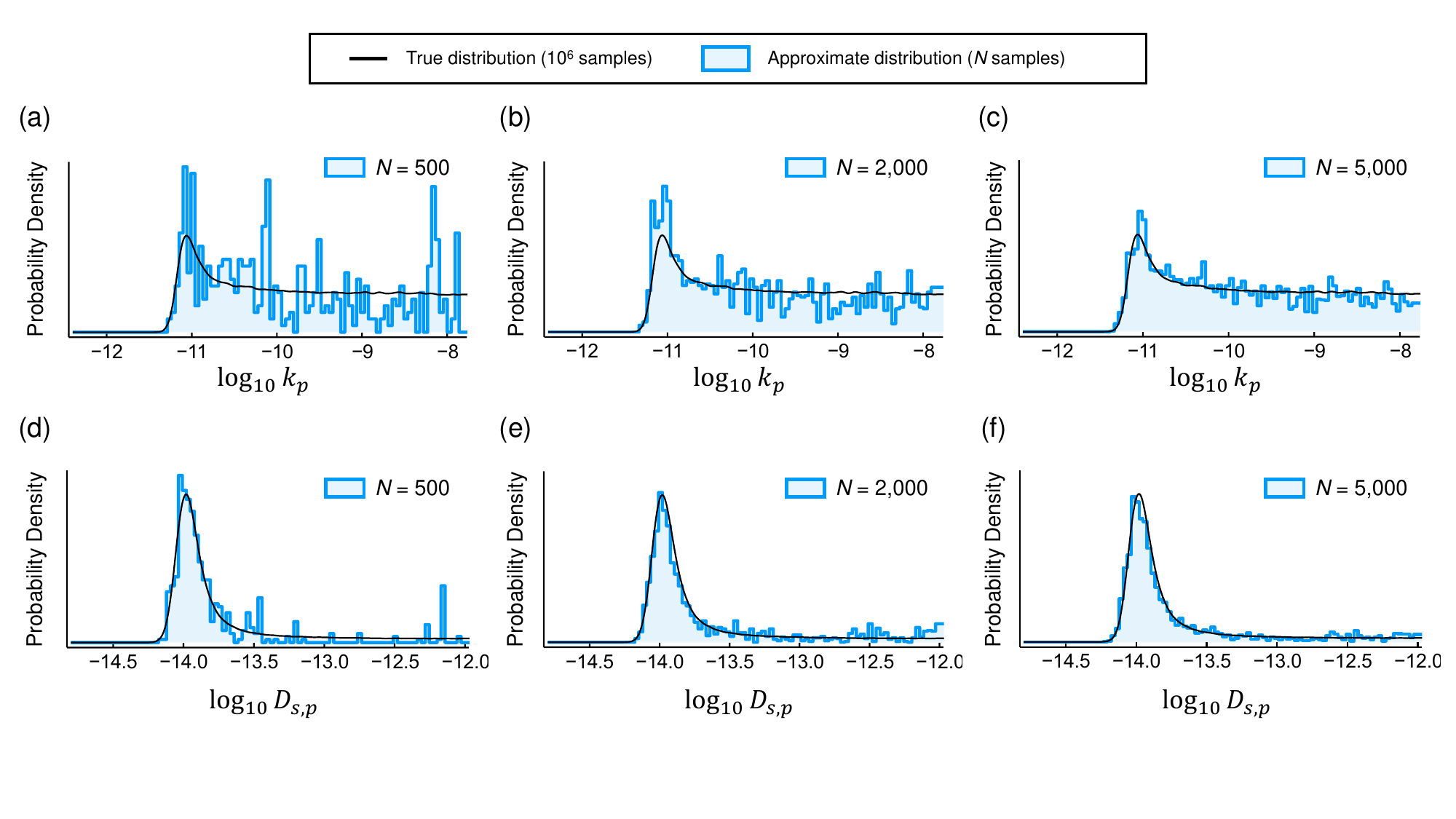}

\vspace{-1.4cm}

\caption{Estimation of the posterior distribution by sampling the parameter space. After a few hundred iterations, the approximate posterior distributions resemble the true distributions with some noise. The true PDF was estimated by sampling the confidence region for 1,000,000 iterations: $k_{p}$ through (a) 500 samples, (b) 2000 samples, (c) 5000 samples, and $D_{s,p}$ through (d) 500 samples, (e) 2000 samples, (f) 5000 samples.}
\label{fig:Posterior}
\end{figure}

As $N_p$ increases, gridding the confidence region becomes prohibitively expensive as the number of model evaluations grows with the order of power $N_p$. The same identifiability trends can be interpreted through MCMC after a sufficient number of iterations which exhibits better scaling. Fig.\ \ref{fig:Posterior} shows the development of the posterior distribution from MCMC after various samples following a burn-in of 500, 2000, and 5000 samples. At 5000 samples (Fig.\ \ref{fig:Posterior}cf), the approximate posterior distributions closely resemble the true posterior distributions (where the true distribution was estimated by running MCMC for 10$^6$ iterations). The same local identifiability trends found with the gridded confidence region can also be interpreted with the posterior distributions from MCMC. Both parameters are practically unidentifiable from the posterior distribution because the distributions plateau towards $+\infty$. The parameters are also locally identifiable because of the prominent peak near their lower bounds.

\begin{figure}[h]
\centering
\includegraphics[width=1\textwidth]{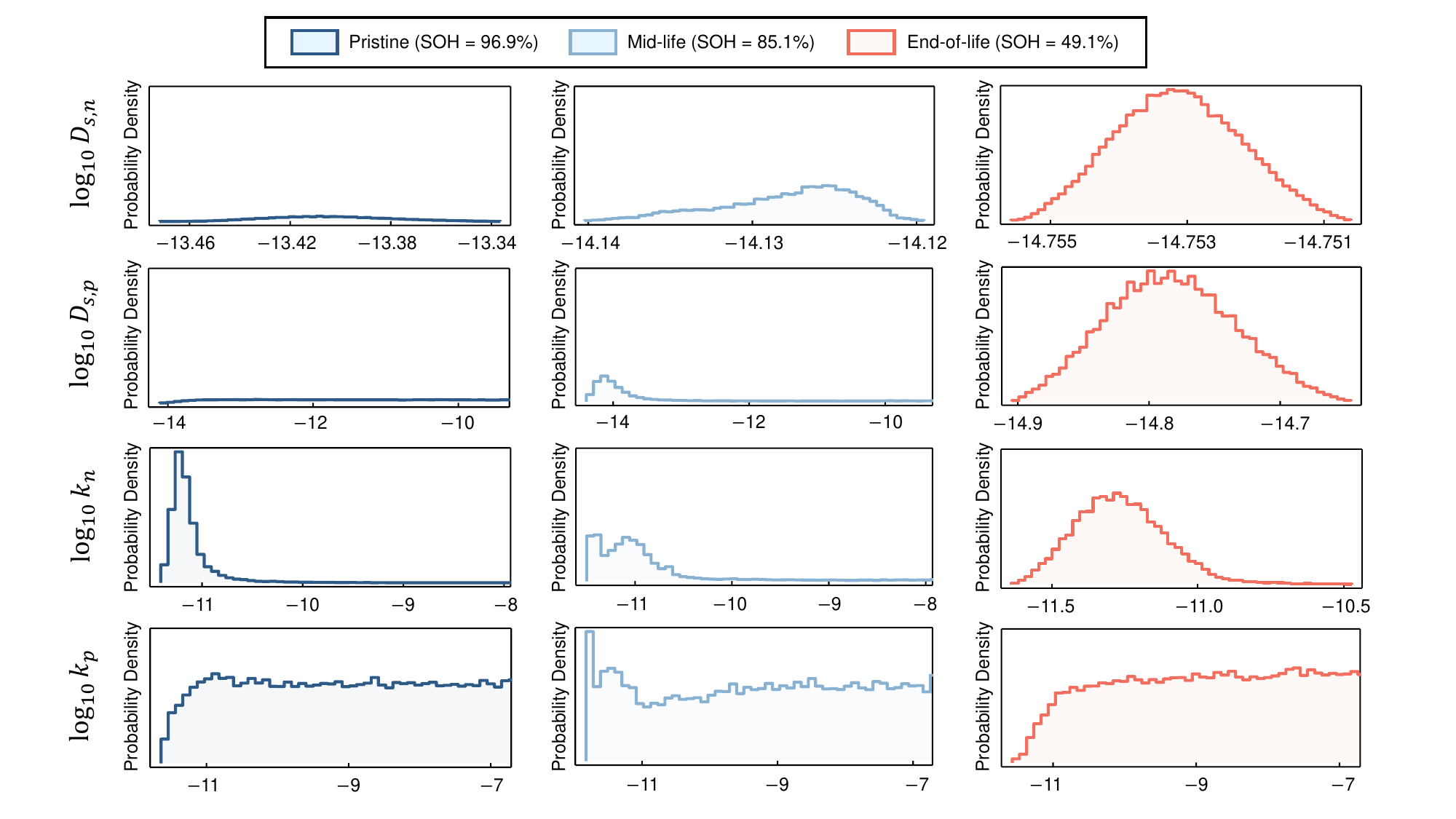}

\vspace{-0.6cm}

\caption{Changing identifiability trends as the cell degrades. For a pristine cell, only $D_{s,n}$ is identifiable. At end-of-life, $D_{s,n}$, $D_{s,p}$, and $k_n$ become identifiable while $k_p$ remains unidentifiable.}
\label{fig:changing_identifiable_trends}
\end{figure}

A parameter identifiability analysis is performed for the diagnostic cycle of every cell using the C/5, 1C, and 2C discharge curves.
These trends change as the battery degrades (Fig.\ \ref{fig:changing_identifiable_trends}). For a pristine cell, $D_{s,n}$ is the only identifiable parameter as its PDF is completely contained within an enclosed region, $k_n$ is practically unidentifiable as its upper bound approaches infinity despite the large peak near the lower bound (i.e., locally identifiable), and both cathode parameters $D_{s,p}$ and $k_p$ are practically unidentifiable. These identifiability trends are consistent with previous articles for LCO discharge curves \cite{berliner2021nonlinear,ramadesigan2011parameter}.
After cycling and degrading the cell, the identifiable parameters are now $D_{s,n}$, $D_{s,p}$, and $k_n$ while $k_p$ remains practically unidentifiable. The mean of each identifiable parameter tends to decrease as a function of the cycle number. The confidence interval of each identifiable parameter tightens as capacity fade increases.

The diffusion and kinetic coefficients are directly related to the movement of Li$^+$ ions in the cell. Large values of $D_{s,i}$ and $k_i$ correspond to faster movement of lithium inside the solid particles and in intercalation, respectively, and small values correspond to slow movement of lithium. The inverse of these coefficients, $1/D_{s,i}$ and $1/k_i$, can be interpreted as \emph{resistances} to lithium flow in these cell sections. Unidentifiable parameters, whose upper bound approaches infinity, have zero resistance, and identifiable parameters have a unique, non-zero resistance. The series resistance of identifiable parameter groupings (e.g., $\left\{ D_{s,p}, k_n \right\}$) have a meaningful and identifiable \emph{total resistance} relationship, e.g.,
\begin{equation}\label{eqn:}
    R_\mathrm{tot} \propto \frac{\alpha}{D_{s,p}} + \frac{\beta}{k_n},
\end{equation}
where $\alpha,\beta$ are positive fitting constants \cite{berliner2021nonlinear}. The changing identifiability trends (Fig.\ \ref{fig:changing_identifiable_trends}) can be interpreted by changing resistances in the cell -- anode resistances are dominant in early life but, as the cell degrades, cathode resistances become measurable (i.e., identifiable). Numerous studies have investigated the loss of active material (LAM) in each electrode during cycling \cite{devie2018intrinsic}. Studies have shown that anode LAM is initially the dominant degradation mechanism, which acts as the limiting electrode. After hundreds of cycles, the cathode LAM can become the primary degradation mechanism, causing a knee-point in the capacity fade curve (Table \ref{tab:map}).

\begin{table}[h]

\caption{Best estimation ($\theta^{*}$) and bounds of $\log_{10} D_{s,n}$, $\log_{10} D_{s,p}$, and $\log_{10} k_{n}$ for a single cell at selected cycles. Practically unidentifiable parameters have upper bounds of $+\infty$.}
\label{tab:map}
\centering

\vspace{+0.2cm}

\resizebox{16cm}{!}{%

\begin{tabular}{cccccccccccc}
\hline
\multirow{2}{*}{C-rate} & \multirow{2}{*}{SOH (\%)} & \multirow{2}{*}{Cycle} & \multicolumn{3}{c}{$\log_{10} D_{s,n}$}                            & \multicolumn{3}{c}{$\log_{10} D_{s,p}$}                            & \multicolumn{3}{c}{$\log_{10} k_{n}$}                             \\ \cmidrule(lr){4-6} \cmidrule(lr){7-9} \cmidrule(lr){10-12}
                        &                      &                        & $\theta^*$ & lower bound & upper bound & $\theta^*$ & lower bound & upper bound & $\theta^*$ & lower bound & upper bound \\ \hline
\multirow{5}{*}{C/5}    & 96.89                & 3                      & -13.4123               & -13.4691    & -13.3386    & -11.7891               & -14.1016    & $+\infty$         & -11.2356               & -11.4084    & -10.7117    \\
                        & 92.03                & 353                    & -13.8416               & -13.8578    & -13.8201    & -13.8258               & -14.2561    & $+\infty$          & -11.0996               & -11.4488    & -10.5916    \\
                        & 85.15                & 773                    & -14.1272               & -14.1349    & -14.1198    & -14.0647               & -14.3874    & $+\infty$           & -11.0947               & -11.4921    & -10.4708    \\
                        & 73.11                & 1193                   & -14.3990               & -14.4027    & -14.3932    & -14.4715               & -14.6339    & -14.1988    & -10.9684               & -11.4155    & -10.2678    \\
                        & 60.44                & 1508                   & -14.5986               & -14.6011    & -14.5959    & -14.6047               & -14.7392    & -14.4277    & -11.1708               & -11.5224    & -10.7465    \\ \hline
\multirow{5}{*}{1C}     & 94.00                & 4                      & -12.9504               & -12.9833    & -12.9032    & -13.2378               & -13.6134    & $+\infty$           & -11.0926               & -11.4292    & -10.7501    \\
                        & 89.02                & 354                    & -13.2537               & -13.2737    & -13.2276    & -9.5828                & -13.7458    & $+\infty$          & -11.6549               & -11.7986    & -10.9207    \\
                        & 80.55                & 774                    & -13.5330               & -13.5443    & -13.5213    & -13.6734               & -13.9970    & $+\infty$           & -11.3667               & -12.0971    & -10.7626    \\
                        & 56.37                & 1194                   & -13.9357               & -13.9441    & -13.9196    & -14.3648               & -14.4676    & -14.2102    & -11.5877               & -12.0410    & -11.2227    \\
                        & 38.12                & 1509                   & -14.2170               & -14.2235    & -14.2116    & -14.4165               & -14.5365    & -14.2574    & -11.8804               & -12.3274    & -11.4914    \\ \hline
\multirow{5}{*}{2C}     & 93.07                & 5                      & -12.5907               & -12.6521    & -12.4817    & -12.7455               & -13.2445    & $+\infty$       & -11.5614               & -11.8172    & -11.2223    \\
                        & 86.38                & 355                    & -13.0125               & -13.0387    & -12.9654    & -13.0800               & -13.4779    & $+\infty$       & -11.8226               & -12.2882    & -11.3668    \\
                        & 76.83                & 775                    & -13.2886               & -13.3021    & -13.2669    & -13.5018               & -13.7450    & -13.0584    & -11.8137               & -12.3387    & -11.3409    \\
                        & 46.50                & 1195                   & -13.7453               & -13.7592    & -13.7078    & -14.0299               & -14.1465    & -13.9412    & -13.1056               & -13.3861    & -12.8286    \\
                        & 24.40                & 1510                   & -14.0743               & -14.1210    & -13.9907    & -14.5462               & -14.6372    & -14.4453    & -13.5577               & -13.8075    & -13.3000    \\ \hline
\end{tabular}
}
\end{table}

\subsection{Degradation diagnosis}

A past study \cite{ramadesigan2011parameter} found that, for the same set of four parameters considered in this article plus electrolyte diffusivity $D$, capacity fade could be predicted for future cycles while only regressing $D_{s,n}$ and $k_n$. Large uncertainties observed for $D_{s,p}$, $D$, and $k_p$ were addressed by fixing their values to be constants, and reductions in the estimated $D_{s,n}$ and $k_n$ with cycle number were observed to follow a power law. Although the approach in \cite{ramadesigan2011parameter} accurately predicted future voltage discharge curves for one cell, the above nonlinear identifiability results indicate that empirical fits only to anode parameters may not apply to a broad range of cells that exhibit competing degradation mechanisms.

\begin{figure}[h]
\centering
\includegraphics[width=1\textwidth]{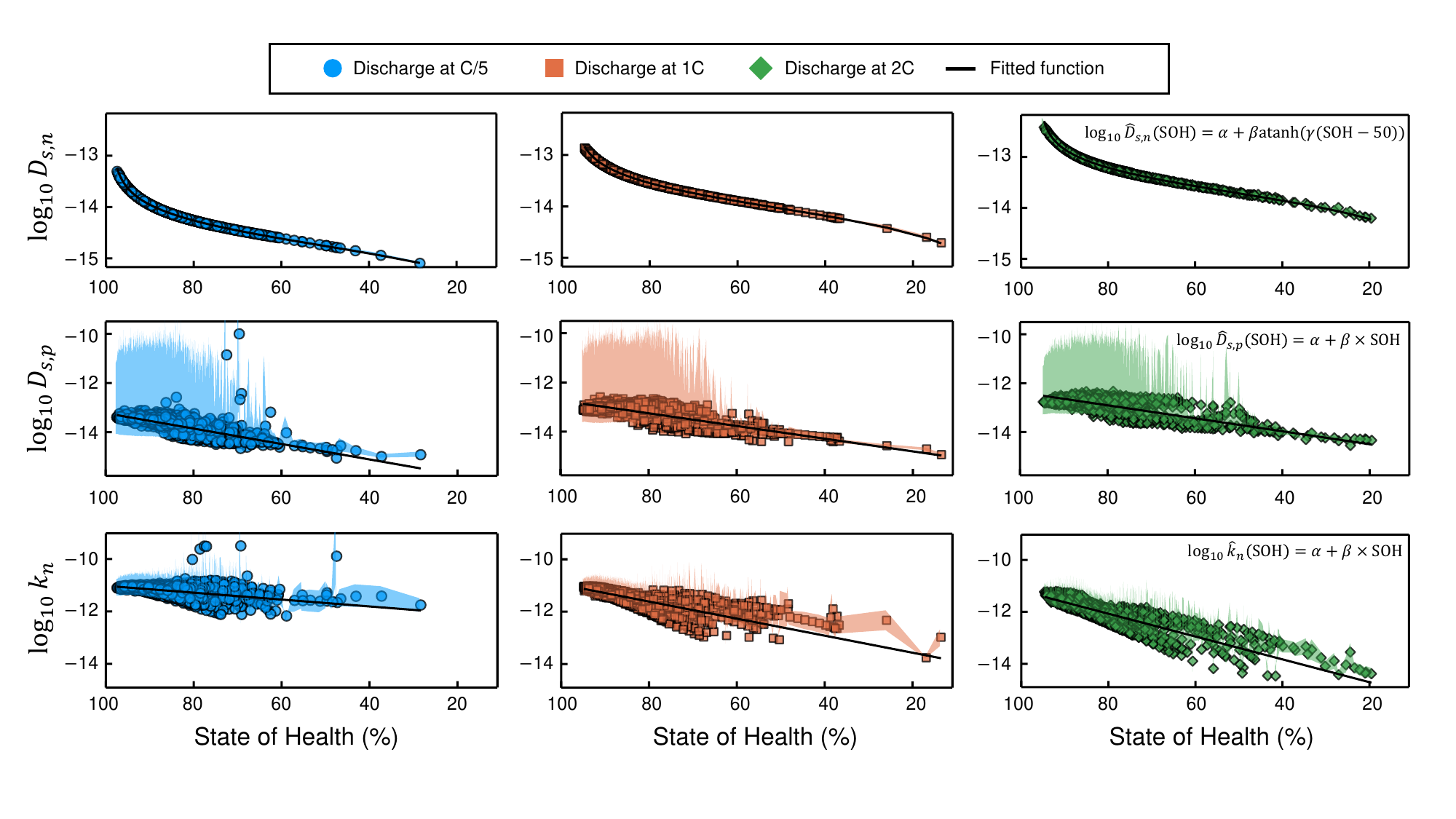}

\vspace{-1cm}

\caption{MCMC results and simple fitting of every cycle for C/5, 1C, and 2C discharge rates of the diagnostic cycles with a weak Gaussian prior for $\log_{10} D_{s,p}$. The highlighted region is the uncertainty for each parameter which is smoothed using an exponentially weighted moving average. Each dot is the logarithm of MAP at a particular SOH for all cells. Fitted parameter equations are generated to describe all MAPs well and are reported in Table \ref{tab:all_cells_fittings}.}
\label{fig:all_cell_fittings}
\end{figure}

At all levels of degradation, $k_p$ is unidentifiable (non-rate limiting) and may be replaced with a sufficiently large constant value ($k_p = 10^{-7}~\text{m}^{5/2} / \text{mol}^{1/2} \text{s}$). At low to moderate levels of degradation, $D_{s,p}$ and $k_n$ are locally identifiable. By adding a weak Gaussian prior to $\log_{10} D_{s,p}$,
\begin{equation}\label{eqn:D_sp_prior}
    P(\log_{10} D_{s,p}) = \mathcal{N}(-15.2, 1^2),
\end{equation}
we shift the median of the posterior distributions of $D_{s,p}$ and $k_n$ towards their lower bounds without greatly affecting the error. Then, we regress the remaining identifiable set, $\{ D_{s,n}, D_{s,p}, k_n \}$, for each cycle. The MCMC results for each cycle (Fig.\ \ref{fig:all_cell_fittings}) are fit to empirical equations using weighted least squares, where the weight is the reciprocal of the variance. $\log_{10} D_{s,n}$ is by far the most well-behaved across all C-rates and levels of degradation with low uncertainty, while $\log_{10} D_{s,p}$ and  $\log_{10} k_n$ have more variation. Starting at large $\text{SOH}$, $\log_{10} D_{s,n}$ quickly slopes downwards before entering a linear regime in $\text{SOH}$. The fit across $\text{SOH}$ is well-approximated with an arctangent function (Table \ref{tab:all_cells_fittings}). The trends for $\log_{10} D_{s,p}$ and  $\log_{10} k_n$ are noisier than the $\log_{10} D_{s,n}$ trend. The difference in uncertainty is likely $\log_{10} D_{s,n}$ encodes the dominant degradation mechanism, while the individual cell-to-cell variation appears in the optimized parameters for $\log_{10} D_{s,p}$ and  $\log_{10} k_n$. 
At large discharge capacities, $\log_{10} D_{s,p}$ has a large level of uncertainty at its upper bound, which is consistent with the identifiability analysis for pristine and mid-life cells (Fig.\ \ref{fig:changing_identifiable_trends}) where $D_{s,p}$ is unidentifiable at low levels of degradation. The mean of $\log_{10} D_{s,p}$ is closer to its lower bound due to the weak prior in \eqref{eqn:D_sp_prior}, which nudges the posterior distribution closer to the lower bound. As the cells reach about 60--70\% SOH, $D_{s,p}$ becomes identifiable.
Broadly, $D_{s,p}$ and  $k_n$ decrease as degradation increases, representing greater internal resistances in the cell. The relationships for $\log_{10} \hat{D}_{s,p}(\text{SOH})$ and  $\log_{10} \hat{k}_n(\text{SOH})$ are approximated with linear fits (Table \ref{tab:all_cells_fittings}). 

\begin{table}[H]
\caption{Fitted equations for the set of diffusion and kinetic parameters as a function of $\text{SOH}$ in units of $\%$ (see Fig.\ \ref{fig:all_cell_fittings}).}
\label{tab:all_cells_fittings}
\centering
\begin{tabular}{llllllcccccccccccccccccccccccccc}
\toprule
	Fitted & Discharge & \multirow{2}[1]{*}{Fitted equation}\\
	parameter, $\hat\theta(\text{SOH})$ & C-rate&\\
	\midrule
                                       & C/5 & $-14.75 + 0.6757 \atanh (2.067(\text{SOH} - 50))$ \\
        $\log_{10} \hat{D}_{s,n}(\text{SOH})$ & 1C  & $-14.04 + 0.6581 \atanh (2.119(\text{SOH} - 50))$ \\
                                       & 2C  & $-13.71 + 0.6368 \atanh (2.190(\text{SOH} - 50))$ \\
        \midrule
                                       & C/5 & $-16.11 + 2.684(\text{SOH})$ \\
        $\log_{10} \hat{D}_{s,p}(\text{SOH})$ & 1C  & $-15.33 + 2.504(\text{SOH})$ \\
                                       & 2C  & $-15.03 + 2.561(\text{SOH})$ \\
        \midrule
                                       & C/5 & $-12.18 + 1.135(\text{SOH})$ \\
        $\log_{10} \hat{k}_n(\text{SOH})$     & 1C  & $-14.36 + 3.409(\text{SOH})$ \\
                                       & 2C  & $-15.68 + 4.502(\text{SOH})$ \\
	\bottomrule
\end{tabular}
\end{table}

\begin{figure}[h]
\centering
\includegraphics[width=1.1\textwidth]{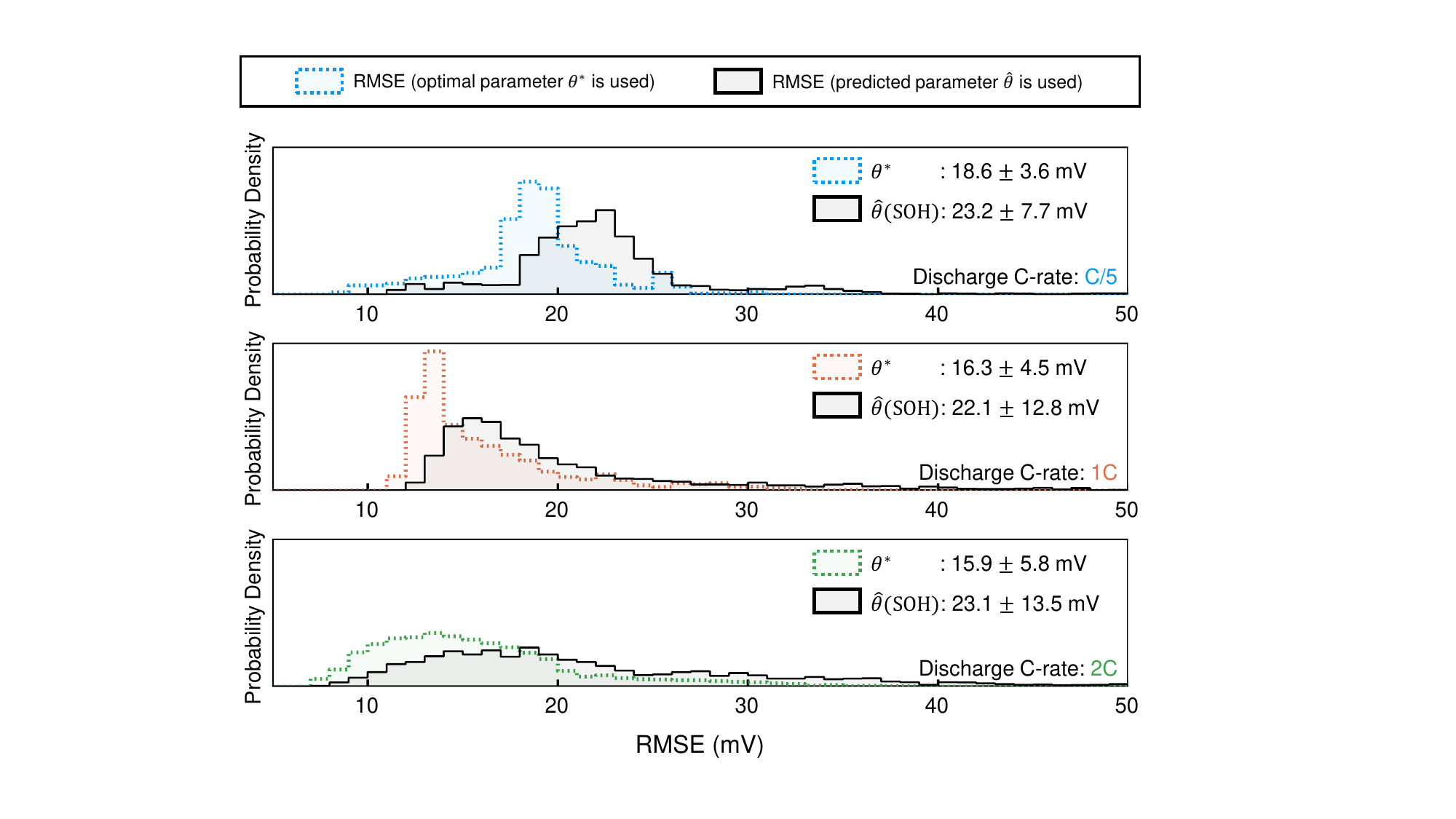}

\vspace{-1cm}

\caption{Histogram of RMSEs from optimizing the parameters for each cycle, $\theta^*$, and with a simple fitted relationship for each parameter, $\hat\theta(\text{SOH})$ (see Table \ref{tab:all_cells_fittings}). The mean and standard deviation are slightly larger using $\hat\theta(\text{SOH})$ compared to $\theta^*$.}
\label{fig:error_histograms}
\end{figure}

The degrees of freedom of optimizing the parameters for every cycle scales with the number of cycles (about 2500) multiplied by the number of regressed parameters (3) -- about 7500 degrees of freedom for each C-rate. In contrast, the degrees of freedom for the fitted parameters are significantly smaller -- only 7 for the three equations for each C-rate (see Table \ref{tab:all_cells_fittings}). It is expected that the sets of optimized parameters, $\theta^*$, will have uniformly smaller errors than those using the fitted parameter relationships, $\hat\theta(\text{SOH})$. Fig.\ \ref{fig:error_histograms} shows the error histograms for the three discharge C-rates with $\theta^*$ and $\hat\theta(\text{SOH})$. On average, the RMSEs increase by 35\% when using $\hat\theta(\text{SOH})$ compared to $\theta^*$, which is acceptable given the significantly smaller degrees of freedom with $\hat\theta(\text{SOH})$. Still, $\hat\theta(\text{SOH})$ is unable to capture significant variation in particular cells and cycles -- the RMSE standard deviations for $\hat\theta(\text{SOH})$ increase by a factor of 2--3 compared to $\theta^*$. The fitted parameters produce greater errors as the capacity fade increases, whereas the optimized parameters produce errors that do not greatly change with capacity fade. One possible explanation is that greater cell-to-cell variation appears as the cells become degraded, leading to greater deviation from the mean as capacity fade increases.

\subsection{Beyond the DFN model}

The DFN model provides a satisfactory coarse-grained description of real data because it includes more physics than the commonly used equivalent circuit models or single-particle models. However, we acknowledge that the DFN model still causes discrepancies with real data because it does not consider a number of physical aspects of the batteries for simplicity:
\begin{enumerate}
\item Staging phase separation in graphite, which leads to non-uniform lithium concentration distributions (in both MPET simulations and experimental imaging~\cite{ferguson2014phase,thomas2017situ,smith2017multiphase}) that bear little resemblance to the assumed shrinking core of the DFN model, except when diffusion dominates at high rates~\cite{fraggedakis2020scaling}. 

\item Lithium plating and SEI growth~\cite{thomas2017situ,gao2021interplay,finegan2020spatial,lu2023multiscale}, the dominant side reactions in graphite anodes, and oxidation-induced cation disordering~\cite{zhuang2022theory,zhuang2023population}, an important degradation mechanism for layered-oxide cathodes, which are only indirectly modeled by re-fitting parameters with aging.

\item Coupled ion-electron transfer (CIET) kinetics of lithium intercalation~\cite{bazant2023unified,fraggedakis2021theory}, which can differ significantly from Butler-Volmer kinetics in the DFN model at extremes values of state of charge (perhaps explaining the larger activation overpotential in experiments compared to DFN simulation near the end of C/5 discharge in Fig.\ \ref{fig:NCA cell data}a).

\item Hybrid porous electrode theory~\cite{liang2023hybrid}, which accounts for significant electrochemical differences between silicon oxide and graphite in the composite anode, leading to nonuniform charging of the two components in both space and time during each cycle. 
\end{enumerate}

Future work could generalize our analysis to capture some of the missing physics using Hybrid MPET, which has recently been applied to similar electrode materials~\cite{liang2023hybrid}. This could lead to different, more realistic values of the model parameters, as well as potentially improved aging predictions, albeit at the cost of greater computational complexity.


\section{Conclusion}\label{sec:conclusions}

\label{conclusion}
In this article, the trajectories of the diffusion coefficient and the reaction rate constant at each electrode over the lifetime are identified via Bayesian inference, and their functional relationship with the SOH is analyzed. A nonlinear identifiability analysis was performed using data across the lifetime of 95 NCA/LiC$_6$-SiO$_{\text{x}}$ cells from a disassembled Tesla Model 3. 7776 diagnostic cycles were evaluated for C/5, 1C, and 2C discharges. Bayesian inference was performed with the DFN model and diffusion/kinetic parameters at each electrode using the MCMC algorithm. Histograms produced from MCMC are used to establish parameter confidence intervals. At low levels of degradation, only the anode solid diffusion coefficient could be uniquely identified from voltage discharge curves, indicating early anode-dominated degradation. At about 60--70\% SOH, the cathode diffusion coefficient and the anode reaction rate constant become identifiable, indicating that anode and cathode degradation pathways become significant and measurable later in life. 

Capacity fade is predicted by empirical models with two or three parameters regressed on the optimal set of parameters, producing average errors of $23\,\text{mV}$. Identifying additional identifiable parameters that significantly contribute to aging and tracking their trajectories could provide insights into lifetime prediction as well as analysis of aging mechanisms. The proposed aging mechanism analysis framework is a versatile approach that can be applied to other battery chemistries.

\section{Code Availability}
The code can be made available upon reasonable request.

\section*{Acknowledgements}
This work was supported by the Toyota Research Institute through the D3BATT Center on Data-Driven Design of Rechargeable Batteries.

\clearpage

\bibliography{apssamp}

\begin{thebibliography}{47}%
\makeatletter
\providecommand \@ifxundefined [1]{%
 \@ifx{#1\undefined}
}%
\providecommand \@ifnum [1]{%
 \ifnum #1\expandafter \@firstoftwo
 \else \expandafter \@secondoftwo
 \fi
}%
\providecommand \@ifx [1]{%
 \ifx #1\expandafter \@firstoftwo
 \else \expandafter \@secondoftwo
 \fi
}%
\providecommand \natexlab [1]{#1}%
\providecommand \enquote  [1]{``#1''}%
\providecommand \bibnamefont  [1]{#1}%
\providecommand \bibfnamefont [1]{#1}%
\providecommand \citenamefont [1]{#1}%
\providecommand \href@noop [0]{\@secondoftwo}%
\providecommand \href [0]{\begingroup \@sanitize@url \@href}%
\providecommand \@href[1]{\@@startlink{#1}\@@href}%
\providecommand \@@href[1]{\endgroup#1\@@endlink}%
\providecommand \@sanitize@url [0]{\catcode `\\12\catcode `\$12\catcode `\&12\catcode `\#12\catcode `\^12\catcode `\_12\catcode `\%12\relax}%
\providecommand \@@startlink[1]{}%
\providecommand \@@endlink[0]{}%
\providecommand \url  [0]{\begingroup\@sanitize@url \@url }%
\providecommand \@url [1]{\endgroup\@href {#1}{\urlprefix }}%
\providecommand \urlprefix  [0]{URL }%
\providecommand \Eprint [0]{\href }%
\providecommand \doibase [0]{http://dx.doi.org/}%
\providecommand \selectlanguage [0]{\@gobble}%
\providecommand \bibinfo  [0]{\@secondoftwo}%
\providecommand \bibfield  [0]{\@secondoftwo}%
\providecommand \translation [1]{[#1]}%
\providecommand \BibitemOpen [0]{}%
\providecommand \bibitemStop [0]{}%
\providecommand \bibitemNoStop [0]{.\EOS\space}%
\providecommand \EOS [0]{\spacefactor3000\relax}%
\providecommand \BibitemShut  [1]{\csname bibitem#1\endcsname}%
\let\auto@bib@innerbib\@empty
\bibitem [{\citenamefont {Newman}\ and\ \citenamefont {Tiedemann}(1975)}]{newman1975porous}%
  \BibitemOpen
  \bibfield  {author} {\bibinfo {author} {\bibfnamefont {J.}~\bibnamefont {Newman}}\ and\ \bibinfo {author} {\bibfnamefont {W.}~\bibnamefont {Tiedemann}},\ }\href@noop {} {\bibfield  {journal} {\bibinfo  {journal} {AIChE Journal}\ }\textbf {\bibinfo {volume} {21}},\ \bibinfo {pages} {25} (\bibinfo {year} {1975})}\BibitemShut {NoStop}%
\bibitem [{\citenamefont {Doyle}\ \emph {et~al.}(1993)\citenamefont {Doyle}, \citenamefont {Fuller},\ and\ \citenamefont {Newman}}]{doyle1993modeling}%
  \BibitemOpen
  \bibfield  {author} {\bibinfo {author} {\bibfnamefont {M.}~\bibnamefont {Doyle}}, \bibinfo {author} {\bibfnamefont {T.~F.}\ \bibnamefont {Fuller}}, \ and\ \bibinfo {author} {\bibfnamefont {J.}~\bibnamefont {Newman}},\ }\href@noop {} {\bibfield  {journal} {\bibinfo  {journal} {Journal of The Electrochemical Society}\ }\textbf {\bibinfo {volume} {140}},\ \bibinfo {pages} {1526} (\bibinfo {year} {1993})}\BibitemShut {NoStop}%
\bibitem [{\citenamefont {Fuller}\ \emph {et~al.}(1994{\natexlab{a}})\citenamefont {Fuller}, \citenamefont {Doyle},\ and\ \citenamefont {Newman}}]{fuller1994relaxation}%
  \BibitemOpen
  \bibfield  {author} {\bibinfo {author} {\bibfnamefont {T.~F.}\ \bibnamefont {Fuller}}, \bibinfo {author} {\bibfnamefont {M.}~\bibnamefont {Doyle}}, \ and\ \bibinfo {author} {\bibfnamefont {J.}~\bibnamefont {Newman}},\ }\href@noop {} {\bibfield  {journal} {\bibinfo  {journal} {Journal of The Electrochemical Society}\ }\textbf {\bibinfo {volume} {141}},\ \bibinfo {pages} {982} (\bibinfo {year} {1994}{\natexlab{a}})}\BibitemShut {NoStop}%
\bibitem [{\citenamefont {Fuller}\ \emph {et~al.}(1994{\natexlab{b}})\citenamefont {Fuller}, \citenamefont {Doyle},\ and\ \citenamefont {Newman}}]{fuller1994simulation}%
  \BibitemOpen
  \bibfield  {author} {\bibinfo {author} {\bibfnamefont {T.~F.}\ \bibnamefont {Fuller}}, \bibinfo {author} {\bibfnamefont {M.}~\bibnamefont {Doyle}}, \ and\ \bibinfo {author} {\bibfnamefont {J.}~\bibnamefont {Newman}},\ }\href@noop {} {\bibfield  {journal} {\bibinfo  {journal} {Journal of The Electrochemical Society}\ }\textbf {\bibinfo {volume} {141}},\ \bibinfo {pages} {1} (\bibinfo {year} {1994}{\natexlab{b}})}\BibitemShut {NoStop}%
\bibitem [{\citenamefont {Galuppini}\ \emph {et~al.}(2023)\citenamefont {Galuppini}, \citenamefont {Berliner}, \citenamefont {Cogswell}, \citenamefont {Zhuang}, \citenamefont {Bazant},\ and\ \citenamefont {Braatz}}]{galuppini2023nonlinear}%
  \BibitemOpen
  \bibfield  {author} {\bibinfo {author} {\bibfnamefont {G.}~\bibnamefont {Galuppini}}, \bibinfo {author} {\bibfnamefont {M.~D.}\ \bibnamefont {Berliner}}, \bibinfo {author} {\bibfnamefont {D.~A.}\ \bibnamefont {Cogswell}}, \bibinfo {author} {\bibfnamefont {D.}~\bibnamefont {Zhuang}}, \bibinfo {author} {\bibfnamefont {M.~Z.}\ \bibnamefont {Bazant}}, \ and\ \bibinfo {author} {\bibfnamefont {R.~D.}\ \bibnamefont {Braatz}},\ }\href@noop {} {\bibfield  {journal} {\bibinfo  {journal} {Journal of Power Sources}\ }\textbf {\bibinfo {volume} {573}},\ \bibinfo {pages} {233009} (\bibinfo {year} {2023})}\BibitemShut {NoStop}%
\bibitem [{\citenamefont {Berliner}\ \emph {et~al.}(2021{\natexlab{a}})\citenamefont {Berliner}, \citenamefont {Zhao}, \citenamefont {Das}, \citenamefont {Forsuelo}, \citenamefont {Jiang}, \citenamefont {Chueh}, \citenamefont {Bazant},\ and\ \citenamefont {Braatz}}]{berliner2021nonlinear}%
  \BibitemOpen
  \bibfield  {author} {\bibinfo {author} {\bibfnamefont {M.~D.}\ \bibnamefont {Berliner}}, \bibinfo {author} {\bibfnamefont {H.}~\bibnamefont {Zhao}}, \bibinfo {author} {\bibfnamefont {S.}~\bibnamefont {Das}}, \bibinfo {author} {\bibfnamefont {M.}~\bibnamefont {Forsuelo}}, \bibinfo {author} {\bibfnamefont {B.}~\bibnamefont {Jiang}}, \bibinfo {author} {\bibfnamefont {W.~H.}\ \bibnamefont {Chueh}}, \bibinfo {author} {\bibfnamefont {M.~Z.}\ \bibnamefont {Bazant}}, \ and\ \bibinfo {author} {\bibfnamefont {R.~D.}\ \bibnamefont {Braatz}},\ }\href@noop {} {\bibfield  {journal} {\bibinfo  {journal} {Journal of The Electrochemical Society}\ }\textbf {\bibinfo {volume} {168}},\ \bibinfo {pages} {090546} (\bibinfo {year} {2021}{\natexlab{a}})}\BibitemShut {NoStop}%
\bibitem [{\citenamefont {Jin}\ \emph {et~al.}(2018)\citenamefont {Jin}, \citenamefont {Danilov}, \citenamefont {Van~den Hof},\ and\ \citenamefont {Donkers}}]{jin2018parameter}%
  \BibitemOpen
  \bibfield  {author} {\bibinfo {author} {\bibfnamefont {N.}~\bibnamefont {Jin}}, \bibinfo {author} {\bibfnamefont {D.~L.}\ \bibnamefont {Danilov}}, \bibinfo {author} {\bibfnamefont {P.~M.~J.}\ \bibnamefont {Van~den Hof}}, \ and\ \bibinfo {author} {\bibfnamefont {M.~C.~F.}\ \bibnamefont {Donkers}},\ }\href@noop {} {\bibfield  {journal} {\bibinfo  {journal} {International Journal of Energy Research}\ }\textbf {\bibinfo {volume} {42}},\ \bibinfo {pages} {2417} (\bibinfo {year} {2018})}\BibitemShut {NoStop}%
\bibitem [{\citenamefont {Jokar}\ \emph {et~al.}(2016)\citenamefont {Jokar}, \citenamefont {Rajabloo}, \citenamefont {D{\'e}silets},\ and\ \citenamefont {Lacroix}}]{jokar2016inverse}%
  \BibitemOpen
  \bibfield  {author} {\bibinfo {author} {\bibfnamefont {A.}~\bibnamefont {Jokar}}, \bibinfo {author} {\bibfnamefont {B.}~\bibnamefont {Rajabloo}}, \bibinfo {author} {\bibfnamefont {M.}~\bibnamefont {D{\'e}silets}}, \ and\ \bibinfo {author} {\bibfnamefont {M.}~\bibnamefont {Lacroix}},\ }\href@noop {} {\bibfield  {journal} {\bibinfo  {journal} {Journal of The Electrochemical Society}\ }\textbf {\bibinfo {volume} {163}},\ \bibinfo {pages} {A2876} (\bibinfo {year} {2016})}\BibitemShut {NoStop}%
\bibitem [{\citenamefont {L{\'{o}}pez~C}\ \emph {et~al.}(2016)\citenamefont {L{\'{o}}pez~C}, \citenamefont {Wozny}, \citenamefont {Flores-Tlacuahuac}, \citenamefont {Vasquez-Medrano},\ and\ \citenamefont {Zavala}}]{lopez2016computational}%
  \BibitemOpen
  \bibfield  {author} {\bibinfo {author} {\bibfnamefont {D.~C.}\ \bibnamefont {L{\'{o}}pez~C}}, \bibinfo {author} {\bibfnamefont {G.}~\bibnamefont {Wozny}}, \bibinfo {author} {\bibfnamefont {A.}~\bibnamefont {Flores-Tlacuahuac}}, \bibinfo {author} {\bibfnamefont {R.}~\bibnamefont {Vasquez-Medrano}}, \ and\ \bibinfo {author} {\bibfnamefont {V.~M.}\ \bibnamefont {Zavala}},\ }\href@noop {} {\bibfield  {journal} {\bibinfo  {journal} {Industrial \& Engineering Chemistry Research}\ }\textbf {\bibinfo {volume} {55}},\ \bibinfo {pages} {3026} (\bibinfo {year} {2016})}\BibitemShut {NoStop}%
\bibitem [{\citenamefont {Forman}\ \emph {et~al.}(2012)\citenamefont {Forman}, \citenamefont {Moura}, \citenamefont {Stein},\ and\ \citenamefont {Fathy}}]{forman2012genetic}%
  \BibitemOpen
  \bibfield  {author} {\bibinfo {author} {\bibfnamefont {J.~C.}\ \bibnamefont {Forman}}, \bibinfo {author} {\bibfnamefont {S.~J.}\ \bibnamefont {Moura}}, \bibinfo {author} {\bibfnamefont {J.~L.}\ \bibnamefont {Stein}}, \ and\ \bibinfo {author} {\bibfnamefont {H.~K.}\ \bibnamefont {Fathy}},\ }\href@noop {} {\bibfield  {journal} {\bibinfo  {journal} {Journal of Power Sources}\ }\textbf {\bibinfo {volume} {210}},\ \bibinfo {pages} {263} (\bibinfo {year} {2012})}\BibitemShut {NoStop}%
\bibitem [{\citenamefont {Aitio}\ \emph {et~al.}(2020)\citenamefont {Aitio}, \citenamefont {Marquis}, \citenamefont {Ascencio},\ and\ \citenamefont {Howey}}]{aitio2020bayesian}%
  \BibitemOpen
  \bibfield  {author} {\bibinfo {author} {\bibfnamefont {A.}~\bibnamefont {Aitio}}, \bibinfo {author} {\bibfnamefont {S.~G.}\ \bibnamefont {Marquis}}, \bibinfo {author} {\bibfnamefont {P.}~\bibnamefont {Ascencio}}, \ and\ \bibinfo {author} {\bibfnamefont {D.}~\bibnamefont {Howey}},\ }\href@noop {} {\bibfield  {journal} {\bibinfo  {journal} {arXiv preprint arXiv:2001.09890}\ } (\bibinfo {year} {2020})}\BibitemShut {NoStop}%
\bibitem [{\citenamefont {Kemper}\ \emph {et~al.}(2015)\citenamefont {Kemper}, \citenamefont {Li},\ and\ \citenamefont {Kum}}]{kemper2015simplification}%
  \BibitemOpen
  \bibfield  {author} {\bibinfo {author} {\bibfnamefont {P.}~\bibnamefont {Kemper}}, \bibinfo {author} {\bibfnamefont {S.~E.}\ \bibnamefont {Li}}, \ and\ \bibinfo {author} {\bibfnamefont {D.}~\bibnamefont {Kum}},\ }\href@noop {} {\bibfield  {journal} {\bibinfo  {journal} {Journal of Power Sources}\ }\textbf {\bibinfo {volume} {286}},\ \bibinfo {pages} {510} (\bibinfo {year} {2015})}\BibitemShut {NoStop}%
\bibitem [{\citenamefont {Ramadesigan}\ \emph {et~al.}(2011)\citenamefont {Ramadesigan}, \citenamefont {Chen}, \citenamefont {Burns}, \citenamefont {Boovaragavan}, \citenamefont {Braatz},\ and\ \citenamefont {Subramanian}}]{ramadesigan2011parameter}%
  \BibitemOpen
  \bibfield  {author} {\bibinfo {author} {\bibfnamefont {V.}~\bibnamefont {Ramadesigan}}, \bibinfo {author} {\bibfnamefont {K.}~\bibnamefont {Chen}}, \bibinfo {author} {\bibfnamefont {N.~A.}\ \bibnamefont {Burns}}, \bibinfo {author} {\bibfnamefont {V.}~\bibnamefont {Boovaragavan}}, \bibinfo {author} {\bibfnamefont {R.~D.}\ \bibnamefont {Braatz}}, \ and\ \bibinfo {author} {\bibfnamefont {V.~R.}\ \bibnamefont {Subramanian}},\ }\href@noop {} {\bibfield  {journal} {\bibinfo  {journal} {Journal of The Electrochemical Society}\ }\textbf {\bibinfo {volume} {158}},\ \bibinfo {pages} {A1048} (\bibinfo {year} {2011})}\BibitemShut {NoStop}%
\bibitem [{\citenamefont {Smith}\ and\ \citenamefont {Bazant}(2017)}]{smith2017multiphase}%
  \BibitemOpen
  \bibfield  {author} {\bibinfo {author} {\bibfnamefont {R.~B.}\ \bibnamefont {Smith}}\ and\ \bibinfo {author} {\bibfnamefont {M.~Z.}\ \bibnamefont {Bazant}},\ }\href@noop {} {\bibfield  {journal} {\bibinfo  {journal} {Journal of The Electrochemical Society}\ }\textbf {\bibinfo {volume} {164}},\ \bibinfo {pages} {E3291} (\bibinfo {year} {2017})}\BibitemShut {NoStop}%
\bibitem [{\citenamefont {Kim}\ \emph {et~al.}(2024)\citenamefont {Kim}, \citenamefont {Schaeffer}, \citenamefont {Berliner}, \citenamefont {Sagnier}, \citenamefont {Bazant}, \citenamefont {Findeisen},\ and\ \citenamefont {Braatz}}]{kim2024fast}%
  \BibitemOpen
  \bibfield  {author} {\bibinfo {author} {\bibfnamefont {M.}~\bibnamefont {Kim}}, \bibinfo {author} {\bibfnamefont {J.}~\bibnamefont {Schaeffer}}, \bibinfo {author} {\bibfnamefont {M.~D.}\ \bibnamefont {Berliner}}, \bibinfo {author} {\bibfnamefont {B.~P.}\ \bibnamefont {Sagnier}}, \bibinfo {author} {\bibfnamefont {M.~Z.}\ \bibnamefont {Bazant}}, \bibinfo {author} {\bibfnamefont {R.}~\bibnamefont {Findeisen}}, \ and\ \bibinfo {author} {\bibfnamefont {R.~D.}\ \bibnamefont {Braatz}},\ }\href@noop {} {\bibfield  {journal} {\bibinfo  {journal} {Journal of The Electrochemical Society}\ }\textbf {\bibinfo {volume} {171}},\ \bibinfo {pages} {090517} (\bibinfo {year} {2024})}\BibitemShut {NoStop}%
\bibitem [{\citenamefont {Jiang}\ \emph {et~al.}(2022)\citenamefont {Jiang}, \citenamefont {Berliner}, \citenamefont {Lai}, \citenamefont {Asinger}, \citenamefont {Zhao}, \citenamefont {Herring}, \citenamefont {Bazant},\ and\ \citenamefont {Braatz}}]{jiang2022fast}%
  \BibitemOpen
  \bibfield  {author} {\bibinfo {author} {\bibfnamefont {B.}~\bibnamefont {Jiang}}, \bibinfo {author} {\bibfnamefont {M.~D.}\ \bibnamefont {Berliner}}, \bibinfo {author} {\bibfnamefont {K.}~\bibnamefont {Lai}}, \bibinfo {author} {\bibfnamefont {P.~A.}\ \bibnamefont {Asinger}}, \bibinfo {author} {\bibfnamefont {H.}~\bibnamefont {Zhao}}, \bibinfo {author} {\bibfnamefont {P.~K.}\ \bibnamefont {Herring}}, \bibinfo {author} {\bibfnamefont {M.~Z.}\ \bibnamefont {Bazant}}, \ and\ \bibinfo {author} {\bibfnamefont {R.~D.}\ \bibnamefont {Braatz}},\ }\href@noop {} {\bibfield  {journal} {\bibinfo  {journal} {Applied Energy}\ }\textbf {\bibinfo {volume} {307}},\ \bibinfo {pages} {118244} (\bibinfo {year} {2022})}\BibitemShut {NoStop}%
\bibitem [{\citenamefont {Sordi}\ \emph {et~al.}(2025)\citenamefont {Sordi}, \citenamefont {Stecchini}, \citenamefont {Evangelista}, \citenamefont {Luder}, \citenamefont {Li}, \citenamefont {Sauer}, \citenamefont {Casalegno},\ and\ \citenamefont {Rabissi}}]{sordi2025degradation}%
  \BibitemOpen
  \bibfield  {author} {\bibinfo {author} {\bibfnamefont {G.}~\bibnamefont {Sordi}}, \bibinfo {author} {\bibfnamefont {A.}~\bibnamefont {Stecchini}}, \bibinfo {author} {\bibfnamefont {R.}~\bibnamefont {Evangelista}}, \bibinfo {author} {\bibfnamefont {D.}~\bibnamefont {Luder}}, \bibinfo {author} {\bibfnamefont {W.}~\bibnamefont {Li}}, \bibinfo {author} {\bibfnamefont {D.}~\bibnamefont {Sauer}}, \bibinfo {author} {\bibfnamefont {A.}~\bibnamefont {Casalegno}}, \ and\ \bibinfo {author} {\bibfnamefont {C.}~\bibnamefont {Rabissi}},\ }\href@noop {} {\bibfield  {journal} {\bibinfo  {journal} {eTransportation}\ }\textbf {\bibinfo {volume} {24}},\ \bibinfo {pages} {100410} (\bibinfo {year} {2025})}\BibitemShut {NoStop}%
\bibitem [{\citenamefont {Liu}\ \emph {et~al.}(2019)\citenamefont {Liu}, \citenamefont {Ivanco},\ and\ \citenamefont {Onori}}]{liu2019aging}%
  \BibitemOpen
  \bibfield  {author} {\bibinfo {author} {\bibfnamefont {Z.}~\bibnamefont {Liu}}, \bibinfo {author} {\bibfnamefont {A.}~\bibnamefont {Ivanco}}, \ and\ \bibinfo {author} {\bibfnamefont {S.}~\bibnamefont {Onori}},\ }\href@noop {} {\bibfield  {journal} {\bibinfo  {journal} {Journal of Energy Storage}\ }\textbf {\bibinfo {volume} {21}},\ \bibinfo {pages} {519} (\bibinfo {year} {2019})}\BibitemShut {NoStop}%
\bibitem [{\citenamefont {Sulzer}\ \emph {et~al.}(2021)\citenamefont {Sulzer}, \citenamefont {Mohtat}, \citenamefont {Aitio}, \citenamefont {Lee}, \citenamefont {Yeh}, \citenamefont {Steinbacher}, \citenamefont {Khan}, \citenamefont {Lee}, \citenamefont {Siegel}, \citenamefont {Stefanopoulou} \emph {et~al.}}]{sulzer2021challenge}%
  \BibitemOpen
  \bibfield  {author} {\bibinfo {author} {\bibfnamefont {V.}~\bibnamefont {Sulzer}}, \bibinfo {author} {\bibfnamefont {P.}~\bibnamefont {Mohtat}}, \bibinfo {author} {\bibfnamefont {A.}~\bibnamefont {Aitio}}, \bibinfo {author} {\bibfnamefont {S.}~\bibnamefont {Lee}}, \bibinfo {author} {\bibfnamefont {Y.~T.}\ \bibnamefont {Yeh}}, \bibinfo {author} {\bibfnamefont {F.}~\bibnamefont {Steinbacher}}, \bibinfo {author} {\bibfnamefont {M.~U.}\ \bibnamefont {Khan}}, \bibinfo {author} {\bibfnamefont {J.~W.}\ \bibnamefont {Lee}}, \bibinfo {author} {\bibfnamefont {J.~B.}\ \bibnamefont {Siegel}}, \bibinfo {author} {\bibfnamefont {A.~G.}\ \bibnamefont {Stefanopoulou}},  \emph {et~al.},\ }\href@noop {} {\bibfield  {journal} {\bibinfo  {journal} {Joule}\ }\textbf {\bibinfo {volume} {5}},\ \bibinfo {pages} {1934} (\bibinfo {year} {2021})}\BibitemShut {NoStop}%
\bibitem [{\citenamefont {Mallarapu}\ \emph {et~al.}(2020)\citenamefont {Mallarapu}, \citenamefont {Kim}, \citenamefont {Carney}, \citenamefont {DuBois},\ and\ \citenamefont {Santhanagopalan}}]{mallarapu2020modeling}%
  \BibitemOpen
  \bibfield  {author} {\bibinfo {author} {\bibfnamefont {A.}~\bibnamefont {Mallarapu}}, \bibinfo {author} {\bibfnamefont {J.}~\bibnamefont {Kim}}, \bibinfo {author} {\bibfnamefont {K.}~\bibnamefont {Carney}}, \bibinfo {author} {\bibfnamefont {P.}~\bibnamefont {DuBois}}, \ and\ \bibinfo {author} {\bibfnamefont {S.}~\bibnamefont {Santhanagopalan}},\ }\href@noop {} {\bibfield  {journal} {\bibinfo  {journal} {ETransportation}\ }\textbf {\bibinfo {volume} {4}},\ \bibinfo {pages} {100065} (\bibinfo {year} {2020})}\BibitemShut {NoStop}%
\bibitem [{\citenamefont {Berliner}\ \emph {et~al.}(2021{\natexlab{b}})\citenamefont {Berliner}, \citenamefont {Cogswell}, \citenamefont {Bazant},\ and\ \citenamefont {Braatz}}]{berliner2021petlion}%
  \BibitemOpen
  \bibfield  {author} {\bibinfo {author} {\bibfnamefont {M.~D.}\ \bibnamefont {Berliner}}, \bibinfo {author} {\bibfnamefont {D.~A.}\ \bibnamefont {Cogswell}}, \bibinfo {author} {\bibfnamefont {M.~Z.}\ \bibnamefont {Bazant}}, \ and\ \bibinfo {author} {\bibfnamefont {R.~D.}\ \bibnamefont {Braatz}},\ }\href@noop {} {\bibfield  {journal} {\bibinfo  {journal} {Journal of The Electrochemical Society}\ }\textbf {\bibinfo {volume} {168}},\ \bibinfo {pages} {090504} (\bibinfo {year} {2021}{\natexlab{b}})}\BibitemShut {NoStop}%
\bibitem [{\citenamefont {Schiesser}(1991)}]{Schiesser}%
  \BibitemOpen
  \bibfield  {author} {\bibinfo {author} {\bibfnamefont {W.~E.}\ \bibnamefont {Schiesser}},\ }\href@noop {} {\emph {\bibinfo {title} {The Numerical Method of Lines: Integration of Partial Differential Equations}}}\ (\bibinfo  {publisher} {Academic Press},\ \bibinfo {address} {San Diego},\ \bibinfo {year} {1991})\BibitemShut {NoStop}%
\bibitem [{\citenamefont {Golberg}(1989)}]{golberg1989genetic}%
  \BibitemOpen
  \bibfield  {author} {\bibinfo {author} {\bibfnamefont {D.~E.}\ \bibnamefont {Golberg}},\ }\href@noop {} {\emph {\bibinfo {title} {Genetic Algorithms in Search, Optimization and Machine Learning}}}\ (\bibinfo  {publisher} {Addison-Wesley},\ \bibinfo {address} {Boston},\ \bibinfo {year} {1989})\BibitemShut {NoStop}%
\bibitem [{\citenamefont {Lee}\ \emph {et~al.}(2024)\citenamefont {Lee}, \citenamefont {Kim}, \citenamefont {Moon},\ and\ \citenamefont {Kim}}]{lee2024multi}%
  \BibitemOpen
  \bibfield  {author} {\bibinfo {author} {\bibfnamefont {E.}~\bibnamefont {Lee}}, \bibinfo {author} {\bibfnamefont {M.}~\bibnamefont {Kim}}, \bibinfo {author} {\bibfnamefont {I.}~\bibnamefont {Moon}}, \ and\ \bibinfo {author} {\bibfnamefont {J.}~\bibnamefont {Kim}},\ }\href@noop {} {\bibfield  {journal} {\bibinfo  {journal} {Chemical Engineering Journal}\ }\textbf {\bibinfo {volume} {490}},\ \bibinfo {pages} {151484} (\bibinfo {year} {2024})}\BibitemShut {NoStop}%
\bibitem [{\citenamefont {Johnson}\ and\ \citenamefont {Wichern}(2002)}]{johnson2002applied}%
  \BibitemOpen
  \bibfield  {author} {\bibinfo {author} {\bibfnamefont {R.~A.}\ \bibnamefont {Johnson}}\ and\ \bibinfo {author} {\bibfnamefont {D.~W.}\ \bibnamefont {Wichern}},\ }\href@noop {} {\emph {\bibinfo {title} {Applied Multivariate Statistical Analysis}}},\ \bibinfo {edition} {5th}\ ed.\ (\bibinfo  {publisher} {Prentice Hall},\ \bibinfo {address} {Upper Saddle River, NJ},\ \bibinfo {year} {2002})\BibitemShut {NoStop}%
\bibitem [{\citenamefont {Meeker}\ and\ \citenamefont {Escobar}(1995)}]{meeker1995teaching}%
  \BibitemOpen
  \bibfield  {author} {\bibinfo {author} {\bibfnamefont {W.~Q.}\ \bibnamefont {Meeker}}\ and\ \bibinfo {author} {\bibfnamefont {L.~A.}\ \bibnamefont {Escobar}},\ }\href@noop {} {\bibfield  {journal} {\bibinfo  {journal} {The American Statistician}\ }\textbf {\bibinfo {volume} {49}},\ \bibinfo {pages} {48} (\bibinfo {year} {1995})}\BibitemShut {NoStop}%
\bibitem [{\citenamefont {Haario}\ \emph {et~al.}(2006)\citenamefont {Haario}, \citenamefont {Laine}, \citenamefont {Mira},\ and\ \citenamefont {Saksman}}]{haario2006dram}%
  \BibitemOpen
  \bibfield  {author} {\bibinfo {author} {\bibfnamefont {H.}~\bibnamefont {Haario}}, \bibinfo {author} {\bibfnamefont {M.}~\bibnamefont {Laine}}, \bibinfo {author} {\bibfnamefont {A.}~\bibnamefont {Mira}}, \ and\ \bibinfo {author} {\bibfnamefont {E.}~\bibnamefont {Saksman}},\ }\href@noop {} {\bibfield  {journal} {\bibinfo  {journal} {Statistics and Computing}\ }\textbf {\bibinfo {volume} {16}},\ \bibinfo {pages} {339} (\bibinfo {year} {2006})}\BibitemShut {NoStop}%
\bibitem [{\citenamefont {Roberts}\ and\ \citenamefont {Rosenthal}(2009)}]{roberts2009examples}%
  \BibitemOpen
  \bibfield  {author} {\bibinfo {author} {\bibfnamefont {G.~O.}\ \bibnamefont {Roberts}}\ and\ \bibinfo {author} {\bibfnamefont {J.~S.}\ \bibnamefont {Rosenthal}},\ }\href@noop {} {\bibfield  {journal} {\bibinfo  {journal} {Journal of Computational and Graphical Statistics}\ }\textbf {\bibinfo {volume} {18}},\ \bibinfo {pages} {349} (\bibinfo {year} {2009})}\BibitemShut {NoStop}%
\bibitem [{\citenamefont {Chib}\ and\ \citenamefont {Greenberg}(1995)}]{chib1995understanding}%
  \BibitemOpen
  \bibfield  {author} {\bibinfo {author} {\bibfnamefont {S.}~\bibnamefont {Chib}}\ and\ \bibinfo {author} {\bibfnamefont {E.}~\bibnamefont {Greenberg}},\ }\href@noop {} {\bibfield  {journal} {\bibinfo  {journal} {The American Statistician}\ }\textbf {\bibinfo {volume} {49}},\ \bibinfo {pages} {327} (\bibinfo {year} {1995})}\BibitemShut {NoStop}%
\bibitem [{\citenamefont {Gering}(2017)}]{gering2017prediction}%
  \BibitemOpen
  \bibfield  {author} {\bibinfo {author} {\bibfnamefont {K.~L.}\ \bibnamefont {Gering}},\ }\href@noop {} {\bibfield  {journal} {\bibinfo  {journal} {Electrochimica Acta}\ }\textbf {\bibinfo {volume} {225}},\ \bibinfo {pages} {175} (\bibinfo {year} {2017})}\BibitemShut {NoStop}%
\bibitem [{\citenamefont {Khuri}\ and\ \citenamefont {Mukhopadhyay}(2010)}]{khuri2010response}%
  \BibitemOpen
  \bibfield  {author} {\bibinfo {author} {\bibfnamefont {A.~I.}\ \bibnamefont {Khuri}}\ and\ \bibinfo {author} {\bibfnamefont {S.}~\bibnamefont {Mukhopadhyay}},\ }\href@noop {} {\bibfield  {journal} {\bibinfo  {journal} {Wiley Interdisciplinary Reviews: Computational Statistics}\ }\textbf {\bibinfo {volume} {2}},\ \bibinfo {pages} {128} (\bibinfo {year} {2010})}\BibitemShut {NoStop}%
\bibitem [{\citenamefont {Moyassari}\ \emph {et~al.}(2021)\citenamefont {Moyassari}, \citenamefont {Streck}, \citenamefont {Paul}, \citenamefont {Trunk}, \citenamefont {Neagu}, \citenamefont {Chang}, \citenamefont {Hou}, \citenamefont {M{\"a}rkisch}, \citenamefont {Gilles},\ and\ \citenamefont {Jossen}}]{moyassari2021impact}%
  \BibitemOpen
  \bibfield  {author} {\bibinfo {author} {\bibfnamefont {E.}~\bibnamefont {Moyassari}}, \bibinfo {author} {\bibfnamefont {L.}~\bibnamefont {Streck}}, \bibinfo {author} {\bibfnamefont {N.}~\bibnamefont {Paul}}, \bibinfo {author} {\bibfnamefont {M.}~\bibnamefont {Trunk}}, \bibinfo {author} {\bibfnamefont {R.}~\bibnamefont {Neagu}}, \bibinfo {author} {\bibfnamefont {C.-C.}\ \bibnamefont {Chang}}, \bibinfo {author} {\bibfnamefont {S.-C.}\ \bibnamefont {Hou}}, \bibinfo {author} {\bibfnamefont {B.}~\bibnamefont {M{\"a}rkisch}}, \bibinfo {author} {\bibfnamefont {R.}~\bibnamefont {Gilles}}, \ and\ \bibinfo {author} {\bibfnamefont {A.}~\bibnamefont {Jossen}},\ }\href@noop {} {\bibfield  {journal} {\bibinfo  {journal} {Journal of The Electrochemical Society}\ }\textbf {\bibinfo {volume} {168}},\ \bibinfo {pages} {020519} (\bibinfo {year} {2021})}\BibitemShut {NoStop}%
\bibitem [{\citenamefont {Lee}\ \emph {et~al.}(2008)\citenamefont {Lee}, \citenamefont {Kim}, \citenamefont {Kim}, \citenamefont {Lim},\ and\ \citenamefont {Lee}}]{lee2008spherical}%
  \BibitemOpen
  \bibfield  {author} {\bibinfo {author} {\bibfnamefont {J.-H.}\ \bibnamefont {Lee}}, \bibinfo {author} {\bibfnamefont {W.-J.}\ \bibnamefont {Kim}}, \bibinfo {author} {\bibfnamefont {J.-Y.}\ \bibnamefont {Kim}}, \bibinfo {author} {\bibfnamefont {S.-H.}\ \bibnamefont {Lim}}, \ and\ \bibinfo {author} {\bibfnamefont {S.-M.}\ \bibnamefont {Lee}},\ }\href@noop {} {\bibfield  {journal} {\bibinfo  {journal} {Journal of Power Sources}\ }\textbf {\bibinfo {volume} {176}},\ \bibinfo {pages} {353} (\bibinfo {year} {2008})}\BibitemShut {NoStop}%
\bibitem [{\citenamefont {van Vlijmen}\ \emph {et~al.}(2023)\citenamefont {van Vlijmen}, \citenamefont {Lam}, \citenamefont {Asinger}, \citenamefont {Cui}, \citenamefont {Ganapathi}, \citenamefont {Sun}, \citenamefont {Herring}, \citenamefont {Gopal}, \citenamefont {Geise}, \citenamefont {Deng} \emph {et~al.}}]{van2023interpretable}%
  \BibitemOpen
  \bibfield  {author} {\bibinfo {author} {\bibfnamefont {B.}~\bibnamefont {van Vlijmen}}, \bibinfo {author} {\bibfnamefont {V.}~\bibnamefont {Lam}}, \bibinfo {author} {\bibfnamefont {P.~A.}\ \bibnamefont {Asinger}}, \bibinfo {author} {\bibfnamefont {X.}~\bibnamefont {Cui}}, \bibinfo {author} {\bibfnamefont {D.}~\bibnamefont {Ganapathi}}, \bibinfo {author} {\bibfnamefont {S.}~\bibnamefont {Sun}}, \bibinfo {author} {\bibfnamefont {P.~K.}\ \bibnamefont {Herring}}, \bibinfo {author} {\bibfnamefont {C.~B.}\ \bibnamefont {Gopal}}, \bibinfo {author} {\bibfnamefont {N.}~\bibnamefont {Geise}}, \bibinfo {author} {\bibfnamefont {H.~D.}\ \bibnamefont {Deng}},  \emph {et~al.},\ }\href@noop {} {\bibfield  {journal} {\bibinfo  {journal} {ChemRxiv}\ } (\bibinfo {year} {2023})}\BibitemShut {NoStop}%
\bibitem [{\citenamefont {Beck}\ and\ \citenamefont {Arnold}(1977)}]{beck1977parameter}%
  \BibitemOpen
  \bibfield  {author} {\bibinfo {author} {\bibfnamefont {J.~V.}\ \bibnamefont {Beck}}\ and\ \bibinfo {author} {\bibfnamefont {K.~J.}\ \bibnamefont {Arnold}},\ }\href@noop {} {\emph {\bibinfo {title} {Parameter Estimation in Engineering and Science}}}\ (\bibinfo  {publisher} {Wiley},\ \bibinfo {address} {New York},\ \bibinfo {year} {1977})\BibitemShut {NoStop}%
\bibitem [{\citenamefont {Devie}\ \emph {et~al.}(2018)\citenamefont {Devie}, \citenamefont {Baure},\ and\ \citenamefont {Dubarry}}]{devie2018intrinsic}%
  \BibitemOpen
  \bibfield  {author} {\bibinfo {author} {\bibfnamefont {A.}~\bibnamefont {Devie}}, \bibinfo {author} {\bibfnamefont {G.}~\bibnamefont {Baure}}, \ and\ \bibinfo {author} {\bibfnamefont {M.}~\bibnamefont {Dubarry}},\ }\href@noop {} {\bibfield  {journal} {\bibinfo  {journal} {Energies}\ }\textbf {\bibinfo {volume} {11}},\ \bibinfo {pages} {1031} (\bibinfo {year} {2018})}\BibitemShut {NoStop}%
\bibitem [{\citenamefont {Ferguson}\ and\ \citenamefont {Bazant}(2014)}]{ferguson2014phase}%
  \BibitemOpen
  \bibfield  {author} {\bibinfo {author} {\bibfnamefont {T.~R.}\ \bibnamefont {Ferguson}}\ and\ \bibinfo {author} {\bibfnamefont {M.~Z.}\ \bibnamefont {Bazant}},\ }\href@noop {} {\bibfield  {journal} {\bibinfo  {journal} {Electrochimica Acta}\ }\textbf {\bibinfo {volume} {146}},\ \bibinfo {pages} {89} (\bibinfo {year} {2014})}\BibitemShut {NoStop}%
\bibitem [{\citenamefont {Thomas-Alyea}\ \emph {et~al.}(2017)\citenamefont {Thomas-Alyea}, \citenamefont {Jung}, \citenamefont {Smith},\ and\ \citenamefont {Bazant}}]{thomas2017situ}%
  \BibitemOpen
  \bibfield  {author} {\bibinfo {author} {\bibfnamefont {K.~E.}\ \bibnamefont {Thomas-Alyea}}, \bibinfo {author} {\bibfnamefont {C.}~\bibnamefont {Jung}}, \bibinfo {author} {\bibfnamefont {R.~B.}\ \bibnamefont {Smith}}, \ and\ \bibinfo {author} {\bibfnamefont {M.~Z.}\ \bibnamefont {Bazant}},\ }\href@noop {} {\bibfield  {journal} {\bibinfo  {journal} {Journal of The Electrochemical Society}\ }\textbf {\bibinfo {volume} {164}},\ \bibinfo {pages} {E3063} (\bibinfo {year} {2017})}\BibitemShut {NoStop}%
\bibitem [{\citenamefont {Fraggedakis}\ \emph {et~al.}(2020)\citenamefont {Fraggedakis}, \citenamefont {Nadkarni}, \citenamefont {Gao}, \citenamefont {Zhou}, \citenamefont {Zhang}, \citenamefont {Han}, \citenamefont {Stephens}, \citenamefont {Shao-Horn},\ and\ \citenamefont {Bazant}}]{fraggedakis2020scaling}%
  \BibitemOpen
  \bibfield  {author} {\bibinfo {author} {\bibfnamefont {D.}~\bibnamefont {Fraggedakis}}, \bibinfo {author} {\bibfnamefont {N.}~\bibnamefont {Nadkarni}}, \bibinfo {author} {\bibfnamefont {T.}~\bibnamefont {Gao}}, \bibinfo {author} {\bibfnamefont {T.}~\bibnamefont {Zhou}}, \bibinfo {author} {\bibfnamefont {Y.}~\bibnamefont {Zhang}}, \bibinfo {author} {\bibfnamefont {Y.}~\bibnamefont {Han}}, \bibinfo {author} {\bibfnamefont {R.~M.}\ \bibnamefont {Stephens}}, \bibinfo {author} {\bibfnamefont {Y.}~\bibnamefont {Shao-Horn}}, \ and\ \bibinfo {author} {\bibfnamefont {M.~Z.}\ \bibnamefont {Bazant}},\ }\href@noop {} {\bibfield  {journal} {\bibinfo  {journal} {Energy \& Environmental Science}\ }\textbf {\bibinfo {volume} {13}},\ \bibinfo {pages} {2142} (\bibinfo {year} {2020})}\BibitemShut {NoStop}%
\bibitem [{\citenamefont {Gao}\ \emph {et~al.}(2021)\citenamefont {Gao}, \citenamefont {Han}, \citenamefont {Fraggedakis}, \citenamefont {Das}, \citenamefont {Zhou}, \citenamefont {Yeh}, \citenamefont {Xu}, \citenamefont {Chueh}, \citenamefont {Li},\ and\ \citenamefont {Bazant}}]{gao2021interplay}%
  \BibitemOpen
  \bibfield  {author} {\bibinfo {author} {\bibfnamefont {T.}~\bibnamefont {Gao}}, \bibinfo {author} {\bibfnamefont {Y.}~\bibnamefont {Han}}, \bibinfo {author} {\bibfnamefont {D.}~\bibnamefont {Fraggedakis}}, \bibinfo {author} {\bibfnamefont {S.}~\bibnamefont {Das}}, \bibinfo {author} {\bibfnamefont {T.}~\bibnamefont {Zhou}}, \bibinfo {author} {\bibfnamefont {C.-N.}\ \bibnamefont {Yeh}}, \bibinfo {author} {\bibfnamefont {S.}~\bibnamefont {Xu}}, \bibinfo {author} {\bibfnamefont {W.~C.}\ \bibnamefont {Chueh}}, \bibinfo {author} {\bibfnamefont {J.}~\bibnamefont {Li}}, \ and\ \bibinfo {author} {\bibfnamefont {M.~Z.}\ \bibnamefont {Bazant}},\ }\href@noop {} {\bibfield  {journal} {\bibinfo  {journal} {Joule}\ }\textbf {\bibinfo {volume} {5}},\ \bibinfo {pages} {393} (\bibinfo {year} {2021})}\BibitemShut {NoStop}%
\bibitem [{\citenamefont {Finegan}\ \emph {et~al.}(2020)\citenamefont {Finegan}, \citenamefont {Quinn}, \citenamefont {Wragg}, \citenamefont {Colclasure}, \citenamefont {Lu}, \citenamefont {Tan}, \citenamefont {Heenan}, \citenamefont {Jervis}, \citenamefont {Brett}, \citenamefont {Das} \emph {et~al.}}]{finegan2020spatial}%
  \BibitemOpen
  \bibfield  {author} {\bibinfo {author} {\bibfnamefont {D.~P.}\ \bibnamefont {Finegan}}, \bibinfo {author} {\bibfnamefont {A.}~\bibnamefont {Quinn}}, \bibinfo {author} {\bibfnamefont {D.~S.}\ \bibnamefont {Wragg}}, \bibinfo {author} {\bibfnamefont {A.~M.}\ \bibnamefont {Colclasure}}, \bibinfo {author} {\bibfnamefont {X.}~\bibnamefont {Lu}}, \bibinfo {author} {\bibfnamefont {C.}~\bibnamefont {Tan}}, \bibinfo {author} {\bibfnamefont {T.~M.}\ \bibnamefont {Heenan}}, \bibinfo {author} {\bibfnamefont {R.}~\bibnamefont {Jervis}}, \bibinfo {author} {\bibfnamefont {D.~J.}\ \bibnamefont {Brett}}, \bibinfo {author} {\bibfnamefont {S.}~\bibnamefont {Das}},  \emph {et~al.},\ }\href@noop {} {\bibfield  {journal} {\bibinfo  {journal} {Energy \& Environmental Science}\ }\textbf {\bibinfo {volume} {13}},\ \bibinfo {pages} {2570} (\bibinfo {year} {2020})}\BibitemShut {NoStop}%
\bibitem [{\citenamefont {Lu}\ \emph {et~al.}(2023)\citenamefont {Lu}, \citenamefont {Lagnoni}, \citenamefont {Bertei}, \citenamefont {Das}, \citenamefont {Owen}, \citenamefont {Li}, \citenamefont {O’Regan}, \citenamefont {Wade}, \citenamefont {Finegan}, \citenamefont {Kendrick} \emph {et~al.}}]{lu2023multiscale}%
  \BibitemOpen
  \bibfield  {author} {\bibinfo {author} {\bibfnamefont {X.}~\bibnamefont {Lu}}, \bibinfo {author} {\bibfnamefont {M.}~\bibnamefont {Lagnoni}}, \bibinfo {author} {\bibfnamefont {A.}~\bibnamefont {Bertei}}, \bibinfo {author} {\bibfnamefont {S.}~\bibnamefont {Das}}, \bibinfo {author} {\bibfnamefont {R.~E.}\ \bibnamefont {Owen}}, \bibinfo {author} {\bibfnamefont {Q.}~\bibnamefont {Li}}, \bibinfo {author} {\bibfnamefont {K.}~\bibnamefont {O’Regan}}, \bibinfo {author} {\bibfnamefont {A.}~\bibnamefont {Wade}}, \bibinfo {author} {\bibfnamefont {D.~P.}\ \bibnamefont {Finegan}}, \bibinfo {author} {\bibfnamefont {E.}~\bibnamefont {Kendrick}},  \emph {et~al.},\ }\href@noop {} {\bibfield  {journal} {\bibinfo  {journal} {Nature Communications}\ }\textbf {\bibinfo {volume} {14}},\ \bibinfo {pages} {5127} (\bibinfo {year} {2023})}\BibitemShut {NoStop}%
\bibitem [{\citenamefont {Zhuang}\ and\ \citenamefont {Bazant}(2022)}]{zhuang2022theory}%
  \BibitemOpen
  \bibfield  {author} {\bibinfo {author} {\bibfnamefont {D.}~\bibnamefont {Zhuang}}\ and\ \bibinfo {author} {\bibfnamefont {M.~Z.}\ \bibnamefont {Bazant}},\ }\href@noop {} {\bibfield  {journal} {\bibinfo  {journal} {Journal of The Electrochemical Society}\ }\textbf {\bibinfo {volume} {169}},\ \bibinfo {pages} {100536} (\bibinfo {year} {2022})}\BibitemShut {NoStop}%
\bibitem [{\citenamefont {Zhuang}\ and\ \citenamefont {Bazant}(2023)}]{zhuang2023population}%
  \BibitemOpen
  \bibfield  {author} {\bibinfo {author} {\bibfnamefont {D.}~\bibnamefont {Zhuang}}\ and\ \bibinfo {author} {\bibfnamefont {M.~Z.}\ \bibnamefont {Bazant}},\ }\href@noop {} {\bibfield  {journal} {\bibinfo  {journal} {Physical Review E}\ }\textbf {\bibinfo {volume} {107}},\ \bibinfo {pages} {044603} (\bibinfo {year} {2023})}\BibitemShut {NoStop}%
\bibitem [{\citenamefont {Bazant}(2023)}]{bazant2023unified}%
  \BibitemOpen
  \bibfield  {author} {\bibinfo {author} {\bibfnamefont {M.~Z.}\ \bibnamefont {Bazant}},\ }\href@noop {} {\bibfield  {journal} {\bibinfo  {journal} {Faraday Discussions}\ }\textbf {\bibinfo {volume} {246}},\ \bibinfo {pages} {60} (\bibinfo {year} {2023})}\BibitemShut {NoStop}%
\bibitem [{\citenamefont {Fraggedakis}\ \emph {et~al.}(2021)\citenamefont {Fraggedakis}, \citenamefont {McEldrew}, \citenamefont {Smith}, \citenamefont {Krishnan}, \citenamefont {Zhang}, \citenamefont {Bai}, \citenamefont {Chueh}, \citenamefont {Shao-Horn},\ and\ \citenamefont {Bazant}}]{fraggedakis2021theory}%
  \BibitemOpen
  \bibfield  {author} {\bibinfo {author} {\bibfnamefont {D.}~\bibnamefont {Fraggedakis}}, \bibinfo {author} {\bibfnamefont {M.}~\bibnamefont {McEldrew}}, \bibinfo {author} {\bibfnamefont {R.~B.}\ \bibnamefont {Smith}}, \bibinfo {author} {\bibfnamefont {Y.}~\bibnamefont {Krishnan}}, \bibinfo {author} {\bibfnamefont {Y.}~\bibnamefont {Zhang}}, \bibinfo {author} {\bibfnamefont {P.}~\bibnamefont {Bai}}, \bibinfo {author} {\bibfnamefont {W.~C.}\ \bibnamefont {Chueh}}, \bibinfo {author} {\bibfnamefont {Y.}~\bibnamefont {Shao-Horn}}, \ and\ \bibinfo {author} {\bibfnamefont {M.~Z.}\ \bibnamefont {Bazant}},\ }\href@noop {} {\bibfield  {journal} {\bibinfo  {journal} {Electrochimica Acta}\ }\textbf {\bibinfo {volume} {367}},\ \bibinfo {pages} {137432} (\bibinfo {year} {2021})}\BibitemShut {NoStop}%
\bibitem [{\citenamefont {Liang}\ and\ \citenamefont {Bazant}(2023)}]{liang2023hybrid}%
  \BibitemOpen
  \bibfield  {author} {\bibinfo {author} {\bibfnamefont {Q.}~\bibnamefont {Liang}}\ and\ \bibinfo {author} {\bibfnamefont {M.~Z.}\ \bibnamefont {Bazant}},\ }\href@noop {} {\bibfield  {journal} {\bibinfo  {journal} {Journal of The Electrochemical Society}\ }\textbf {\bibinfo {volume} {170}},\ \bibinfo {pages} {093510} (\bibinfo {year} {2023})}\BibitemShut {NoStop}%
\end{thebibliography}%

\end{document}